\documentclass[apj,onecolumn]{emulateapj}
\usepackage{apjfonts}

\shorttitle{QSO LFs from Joint Evolution of BHs and Host Galaxies}
\shortauthors{Lapi et al.} \journalinfo{Accepted on ApJ.}

\begin{document}

\title{Quasar Luminosity Functions from Joint Evolution
of Black Holes and Host Galaxies}

\author{A. Lapi\altaffilmark{1}, F. Shankar\altaffilmark{3}, J.
Mao\altaffilmark{1}, G.L. Granato\altaffilmark{2,1}, L.
Silva\altaffilmark{4}, G. De Zotti\altaffilmark{2,1}, L.
Danese\altaffilmark{1,2}}\altaffiltext{1}{Astrophysics Sector,
SISSA/ISAS, Via Beirut 2-4, I-34014 Trieste, Italy.}
\altaffiltext{2}{INAF, Osservatorio Astronomico di Padova, Vicolo
dell' Osservatorio 5, I-35122 Padova, Italy.} \altaffiltext{3}{Dept.
of Astronomy, Ohio State University, 140 W $18^{\mathrm{th}}$
Avenue, OH 43210-1173 Columbus, USA.}\altaffiltext{4}{INAF,
Osservatorio Astronomico di Trieste, Via Tiepolo 11, I-34131
Trieste, Italy.}

\begin{abstract}
We show that our previously proposed anti-hierarchical baryon
collapse scenario for the joint evolution of black holes and host
galaxies predicts quasar luminosity functions at redshifts $1.5\la
z\la 6$ and local properties in nice agreement with observations. In
our model the quasar activity marks and originates the transition
between an earlier phase of violent and heavily dust-enshrouded
starburst activity promoting rapid black hole growth, and a later
phase of almost passive evolution; the former is traced by the
submillimeter-selected sources, while the latter accounts for the
high number density of massive galaxies at substantial redshifts
$z\ga 1.5$, the population of Extremely Red Objects, and the
properties of local ellipticals.
\end{abstract}


\section{Introduction}

Observations in the optical and in the X-ray bands have shown since
a couple of decades that very bright quasars (QSOs) occur very soon
in the history of the Universe, and that their average luminosity
significantly declines from redshift $z\sim 2$ toward the present
(e.g., Mathez 1978; Giacconi 1985; Schmidt \& Green 1986). Modern
observations have confirmed and detailed such a picture (e.g., Fan
et al. 2004, 2006; Richards et al. 2005, 2006; Cristiani et al.
2004; Barger et al. 2005; Tozzi et al. 2001; Brandt \& Hasinger
2005), measuring QSO luminosity functions (LFs) up to  $z\sim 6$.

Supermassive black holes (BHs) were found to be ubiquitous in the
centers of spheroidal galaxies (Kormendy \& Richstone 1995).
Moreover, a narrow relationship between the central supermassive BH
mass and the stellar mass/luminosity of the spheroidal/bulge
component has been firmly established by observations (e.g.,
Magorrian et al. 1998; Marconi \& Hunt 2003). The BH mass also
strictly correlates with the velocity dispersion of the spheroidal
component (e.g., Ferrarese \& Merritt 2000; Gebhardt et al. 2000;
Tremaine et al. 2002). Contrariwise, it has been shown that the BH
mass is on average at least $10^2$ times smaller in late-type
irregular and/or spiral galaxies than in ellipticals with the same
luminosity (see Salucci et al. 2000). All together these
observations witness the strict connection between the mass of the
BH and that of the old stars formed at $z\ga 1$ within the host
galaxy. Recently, the studies of QSO host galaxies have been
extended to high redshift (e.g., Dunlop et al. 2003; Floyd et al.
2004), with the general result that they are typically early-types.

The stellar populations of ellipticals are old and have similar
ages/formation histories, although there is some trend for more
massive systems to be on average older and more metal-rich (Sandage
\& Visvanathan 1978; Bernardi et al. 1998; Trager et al. 2000;
Terlevich \& Forbes 2002; Thomas et al. 2005). A color-magnitude
relation is well established, indicating that more luminous
spheroids are redder (Bower et al. 1992). The widely accepted
interpretation is that brighter objects are richer in metals and
that the spread of their star formation epochs is small enough to
avoid smearing of their colors. In fact, the slope of the relation
seems not to change with redshift (Ellis et al. 1997; Kodama et al.
1998), supporting this view. Note, however, that the trend of having
on average older stellar populations in more massive ellipticals may
also constitute an important ingredient (e.g., J{\o}rgensen et al.
1999; Caldwell, Rose \& Concannon 2003; Nelan et al. 2005).

The star formation history of spheroidal galaxies is mirrored in the
Fundamental Plane (e.g., Jedrzejewski et al. 1987; Dressler et al.
1987) and in its evolution with redshift. Ellipticals adhere to this
plane with a surprisingly low orthogonal scatter (around $15\%$), as
expected for a homogeneous family. Furthermore, recent studies
(e.g., Treu et al. 2002; van der Wel et al. 2004; Holden et al.
2005) suggest that ellipticals, both in the field and in the
clusters, lay on the fundamental plane up to $z\approx 1$,
consistent with massive spheroids being old and quiescent.
Progenitors of local massive early-type galaxies have been
identified through K-band and Spitzer surveys at substantial
redshifts, $z\ga 1$. Direct evidence that massive galaxies with
stellar content $M_{\star} \gtrsim 10^{11}\, M_{\odot}$ were in
place at $z \gtrsim 2$ is provided by recent K-band surveys (e.g.,
Cimatti et al. 2002; Kashikawa et al. 2003; Fontana et al. 2004;
Bundy, Ellis \& Conselice 2005). The space density of Extremely Red
Objects (EROs) at $z\gtrsim 3$ is only a factor about $5-10$ less
than that at $z\sim 1$ (e.g., Tecza et al. 2004). This implies a
phase of extremely high star formation rate (SFR), from several
hundreds to thousands solar masses per year, taking place at high
$z$. Such a phase is also witnessed by the sub-mm galaxy counts
(e.g., Chapman et al. 2003; 2005).

On the theoretical side, the hierarchical clustering paradigm led to
the development of various semi-analytic models for galaxy
formation. These generally share the basic assumption that gravity
is the main driver in shaping the structure and morphology of
galaxies. In this scenario, the gas cools and form stars following
the collapse of dark matter (DM) halos. In the standard cosmology,
small DM objects form first and merge together to make larger ones.
This scenario then implies that large ellipticals form late, by
mergers of disk/bulge systems made primarily of stars, recently
indicated as `dry mergers' (e.g., Naab et al. 2006).

But the early appearance of QSOs with high luminosity and presumably
huge BH mass is at variance with expectations from this framework
(e.g., Bromley et al. 2004). This apparent contradiction may be
solved by invoking a higher efficiency in forming massive BHs in
smaller galaxy halos at higher redshift (e.g., Haehnelt \& Rees
1993; Haiman \& Loeb 1998; Wyithe \& Loeb 2003; Mahmood et al.
2005).

Granato et al. (2001, 2004), instead, explored the possibility of
reconciling the observed `downsizing' of QSOs and spheroidal
galaxies with the bottom-up DM hierarchy (see also Somerville et al.
2004; Baugh et al. 2005; Croton et al. 2006; Hopkins et al. 2006).
This has been done by developing a simple model that incorporates
the main physical aspects of the DM and baryons residing within
galactic halos. This new approach emphasized the role of energy
feedback processes both from supernovae (SNae), capable of unbinding
the gas in low-mass systems, and from the QSO phase of the
supermassive BHs, capable of ejecting gas from the largest objects
(see Di Matteo, Springel \& Hernquist 2005; Lapi, Cavaliere \& Menci
2005). These feedback processes can actually reverse the formation
sequence of visible galaxies with respect to that of DM halos. Hence
large galaxies end their star formation, and have their BHs shining
as QSOs early on. On the contrary, the star formation and the QSO
phase are more prolonged in smaller halos (hence the name of
`Anti-hierarchical Baryon Collapse', or ABC, scenario).

While in previous works we focused on reproducing the properties of
spheroidal galaxies, in this paper we explore the constraints on
model parameters set by the redshift-dependent LFs of QSOs. Our plan
is as follows: in \S~2 we briefly recall the main features of the
model by Granato et al. (2004); in \S~3 we describe the procedure
adopted to compute the QSO LFs, the supermassive BH mass function,
and other galactic observables; \S~4 is devoted to the illustration
of our results; in \S~5 we make a critical comparison of our
findings with other models in the literature; finally, in \S~6 we
summarize and discuss our conclusions.

Throughout this work we adopt the cosmology indicated by the WMAP
data (Bennett et al. 2003; Spergel et al. 2006), i.e., a flat
Universe with matter density $\Omega_M\approx 0.27$, baryon density
$\Omega_b\approx 0.044$ and Hubble constant $H_0\approx 71$ km
s$^{-1}$ Mpc$^{-1}$.

\section{The model from the ground up}

This paper is based on the semi-analytic model developed by Granato
et al. (2004), which follows the evolution of baryons within
proto-spheroids through simple and physically grounded recipes.

We defer the interested reader to that paper for a full account of
the physical justification and a detailed description of the model.
Here we provide a short summary of its main features, focusing on
the aspects relevant to our discussion on QSO LFs and BH
demographics.

\subsection{The dark matter sector}

As for the treatment of the DM in galaxies, the model follows the
standard hierarchical clustering framework, taking also into account
the results by Wechsler et al. (2002), and Zhao et al. (2003a;
2003b). Their simulations have shown that the growth of a halo
occurs in two different phases: a first regime of fast accretion in
which the potential well is built up by the sudden mergers of many
clumps with comparable masses; and a second regime of slow accretion
in which mass is added in the outskirts of the halo, without
affecting the central region where the galactic structure resides.

This means that the halos harboring a massive elliptical galaxy once
created, even at high redshift, are rarely destroyed. At low
redshifts they are incorporated within groups and clusters of
galaxies. Support to this view comes from studies of the mass
structure of elliptical galaxies, which are found not to show strong
signs of evolution since redshift $z\approx 1$ (Koopmans et al.
2006). Note that, as pointed out in \S~1, the BH mass is strictly
correlated with properties (mass and velocity dispersion) of the old
stars in massive early-type galaxies, formed at least $8$ Gyr ago
(see Thomas et al. 2005) and, as a consequence, in massive galaxy
halos virialized at $z\ga 1.5$.

We deal just with the latter redshift range ($z \ga 1.5$), where a
good approximation of the halo formation rates is provided by the
positive term in the cosmic time derivative of the cosmological mass
function (e.g., Haehnelt \& Rees 1993; Sasaki 1994). For DM halos
with mass $M_{\mathrm{vir}}$ at time $t_{\mathrm{vir}}$, these
formation rates are given by
\begin{equation}
{\mathrm{d}^2\, N_{\mathrm{ST}}\over \mathrm{d} t_{\mathrm{vir}}\,
\mathrm{d} M_{\mathrm{vir}}}=\left[{a\,
\delta_c(t_{\mathrm{vir}})\over \sigma^2(M_{\mathrm{vir}})}+{2\,
p\over \delta_c(t_{\mathrm{vir}})}\,
{\sigma^{2\,p}(M_{\mathrm{vir}})\over
\sigma^{2\,p}(M_{\mathrm{vir}})+ a^p\,
\delta_c^{2\,p}(t_{\mathrm{vir}})}\right]\,
\left|{\mathrm{d}\delta_c\over \mathrm{d} t_{\mathrm{vir}}}\right|\,
N_{\mathrm{ST}}(M_{\mathrm{vir}}, t_{\mathrm{vir}})~~,
\label{eq|DMrates}
\end{equation}
where $N_{ST}(M_{\mathrm{vir}},t)$ is the Sheth \& Tormen (1999,
2002) version of the PS mass function (Press \& Schecter 1974). In
the above equation, $a=0.707$ and $p=0.3$ are constants obtained
from comparison of the mass function with the outcome of $N$-body
simulations; $\sigma(M_{\mathrm{vir}})$ is the mass variance of the
primordial perturbation field, computed from the Bardeen et al.
(1986) power spectrum with correction for baryons (Sugiyama 1995),
and normalized to $\sigma_8\approx 0.84$ on a scale of $8\,h^{-1}$
Mpc; $\delta_c(t_{\mathrm{vir}})$ is the critical threshold for
collapse, extrapolated from the linear perturbation theory. Note
that our adoption of the Sheth \& Tormen mass function to construct
the rates is mandatory, since it is well-known that the canonical
Press \& Schechter theory strongly underpredicts the number of
massive halos, particularly at the high redshifts relevant for the
computation of the QSO LFs.

We confine our analysis to galaxy halo masses
$M_{\mathrm{vir}}^{\mathrm{min}}\ga 2\times 10^{11}\, M_{\odot}$,
since we are interested in following the evolutionary history of
bright QSOs and of their host galaxies. At the other end, weak
lensing observations (e.g., Kochanek \& White 2001; Kleinheinrich et
al. 2004) and kinematical measurements (e.g., Kronawitter et al.
2000; Gerhard et al. 2001) suggest an upper limit in galaxy halo
mass $M_{\mathrm{vir}}^{\mathrm{max}}\approx 2\times 10^{13}\,
M_{\odot}$. The same limit is also indicated by the theoretical
analysis of Cirasuolo et al. (2005) on the velocity dispersion
function of early-type galaxies. These limits in mass and redshift
ensure that the positive cosmic time derivative of the halo mass
function is a good approximation to the formation rate of DM halos,
as the negative term is negligible.

Note, however, that since we are interested in DM halos associated
to a single galaxy, we need to correct the Sheth \& Tormen halo mass
function by: (i) accounting for the possibility that a DM halo
contains various sub-halos each hosting a galaxy; (ii) removing
halos corresponding to galaxy systems rather than to individual
galaxies. We deal with (ii) by simply cutting-off the halo mass
function at $M_{\mathrm{vir}}^{\mathrm{max}}\approx 2\times
10^{13}\,M_{\odot}$, beyond which the probability of having multiple
galaxies within a halo quickly becomes very high (e.g.,
Magliocchetti \& Porciani 2003). As for (i), we add the sub-halo
mass function, following the procedure described Vale \& Ostriker
(2004; 2006) and Shankar et al. (2006), and using the fit to the
sub-halo mass function at various redshifts provided by van den
Bosch, Tormen \& Giocoli (2005). However we have checked that for
the masses and redshifts relevant here ($z\sim 1.5-6$ and
$M_{\mathrm{vir}}\sim 2\times 10^{11}\, M_{\odot}-2\times 10^{13}\,
M_{\odot}$), the total mass function (sub-halo $+$ halo) differs
from the halo mass function by less than $5\%$. Thus, using simply
the halo mass function to derive the DM halo formation rates is a
good approximation.

\begin{figure}[t]
\epsscale{.8}\plottwo{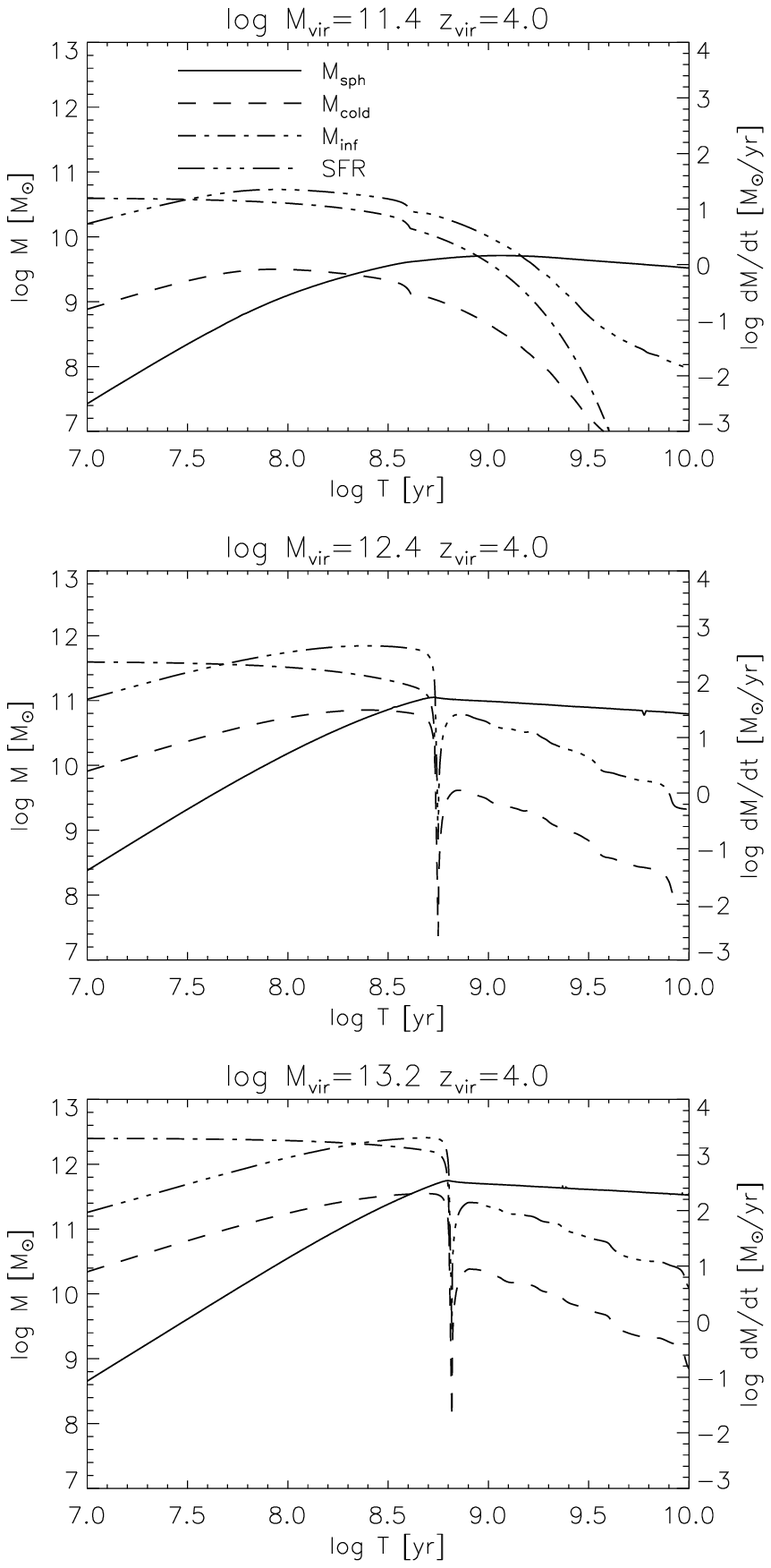}{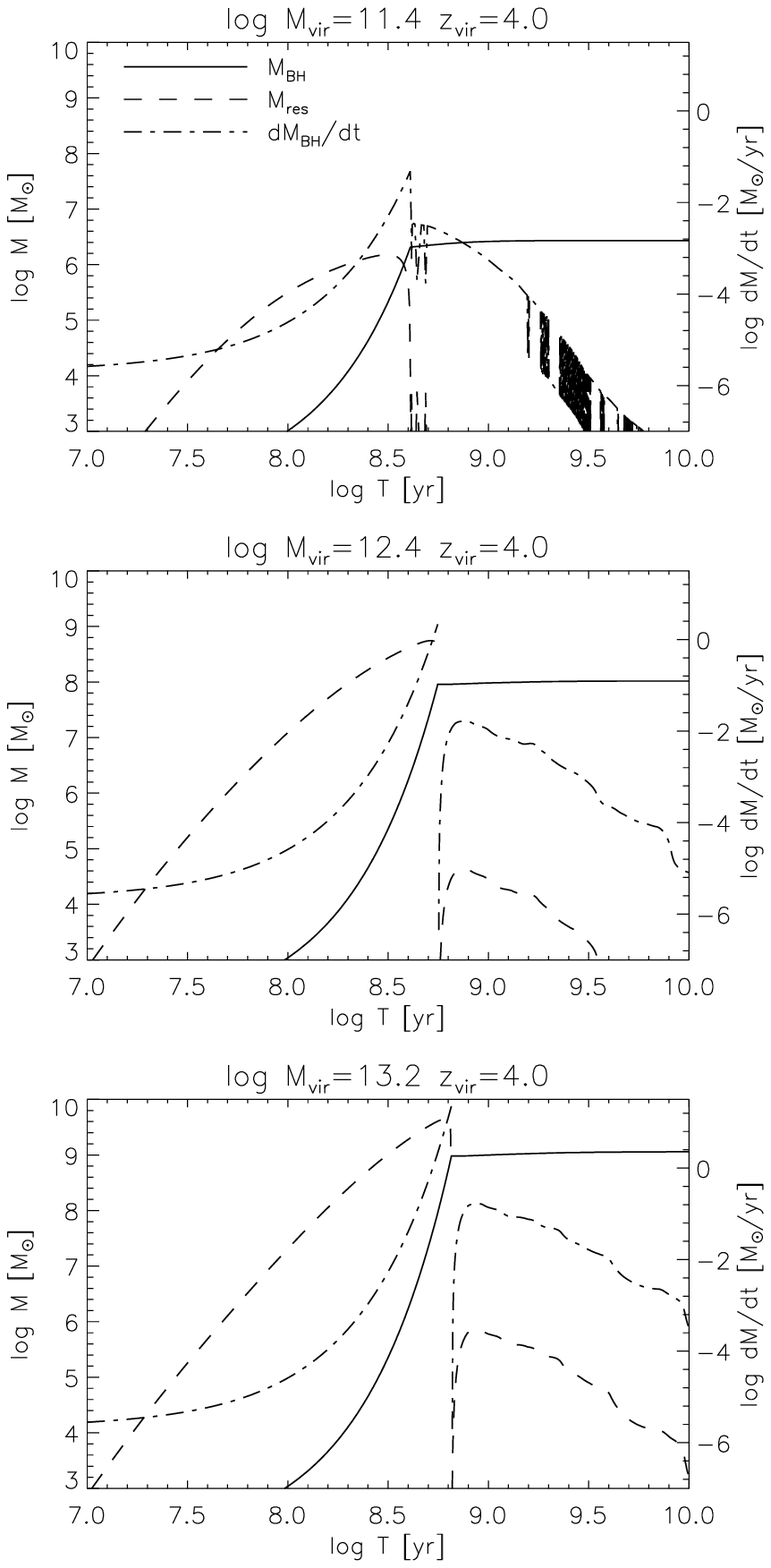}\caption{Left
panel: dependence on galactic age of the stellar mass, infalling
mass, cold gas mass, and star formation rate within halos of various
masses, virialized at redshift $4$. Right panel: same for the BH
mass, reservoir mass, and BH accretion rates.}\label{fig|star&BH}
\end{figure}

\subsection{The baryonic sector}

The physics governing the evolution of the baryons is much more
complex than for the DM. The main features of our model can be
summarized as follows (see Granato et al. 2004 and Cirasuolo et al.
2005 for additional details).

During or soon after the formation of the host DM halo, the baryons
falling into the newly created potential well are shock-heated to
the virial temperature. The hot gas is (moderately) clumpy and cools
quickly especially in the denser central regions, triggering a
strong burst of star formation. The radiation drag due to starlight
acts on the gas clouds, reducing their angular momentum. As a
consequence, a fraction of the cool gas can fall into a reservoir
around the central supermassive BH, and eventually accretes onto it
by viscous dissipation, powering the nuclear activity. The energy
fed back to the gas by SN explosions and BH activity regulates the
ongoing star formation and the BH growth. Eventually, most of the
gas is unbound from the DM potential well, so that the star
formation and the BH activity come to an end on a timescale shorter
for the more massive galaxies.

In Appendix A and Table~A1 we present the basic equations and
parameters controlling the evolution of the baryonic component in
our model, once the halo mass and the virialization redshift are
given. These equations can be numerically integrated to yield, among
others, the SFR and the accretion rate onto the central BH as
function of cosmic time. In Fig.~\ref{fig|star&BH} we plot the basic
outputs of the model.

Initially, the cooling is rapid and the star formation is very high;
thus the radiation drag is efficient in accumulating mass into the
reservoir. The BH starts growing from an initial seed with mass
$10^2\, M_{\odot}$ already in place at the galactic center. Since
there is plenty of material in this phase, the accretion is
Eddington (or moderately super-Eddington) limited (e.g., Small \&
Blandford 1992; Blandford 2004). This regime goes on until the
energy feedback from the BH is strong enough to unbind the gas from
the potential well, a condition occurring around the peak of the
accretion curve. Subsequently, the SFR drops substantially, the
radiation drag becomes inefficient, the storage of matter in the
reservoir and the accretion onto the BH decrease by a large factor.
The drop is very pronounced for massive halos $M_{\mathrm{vir}}\ga
10^{12}\, M_{\odot}$, while for smaller masses a smoother declining
phase can continue for several Gyrs, and the BH and stellar masses
can further increase by a substantial factor.

Before the peak, radiation is highly obscured by the surrounding
dust. In fact, these proto-galaxies are extremely faint in the
UV-optical rest frame and are more easily selected at (sub-)mm
wavelengths. The nuclear emission is heavily obscured too; however,
since the absorption significantly decreases with increasing X-ray
energy of photons, it is easier to detect it in hard X-ray bands. On
the other hand, in the proximity of the accretion peak, i.e., when
the central supermassive BH is massive and powerful enough to remove
most of the gas and dust from the surroundings, the active nucleus
shines as an optical QSO.

The list of the observations well fitted by the model is presented
in Table~A2, along with a brief description of the main underlying
assumptions. In order to translate the SFR and the mass in stars
into more directly observable quantities such as broad-band
luminosities, including dust extinction, we exploit the GRASIL code
(Silva et al. 1998, see \url{http://web.pd.astro.it/granato}). As
for the QSOs, the model yields bolometric luminosities that are
translated into luminosities in a given band by means of appropriate
bolometric corrections. The epoch-dependent luminosity functions are
derived combining the results obtained for halos of fixed mass and
virialization redshift with the halo formation rates presented in
\S~2.1.

\begin{figure}[t]
\epsscale{1.}\plottwo{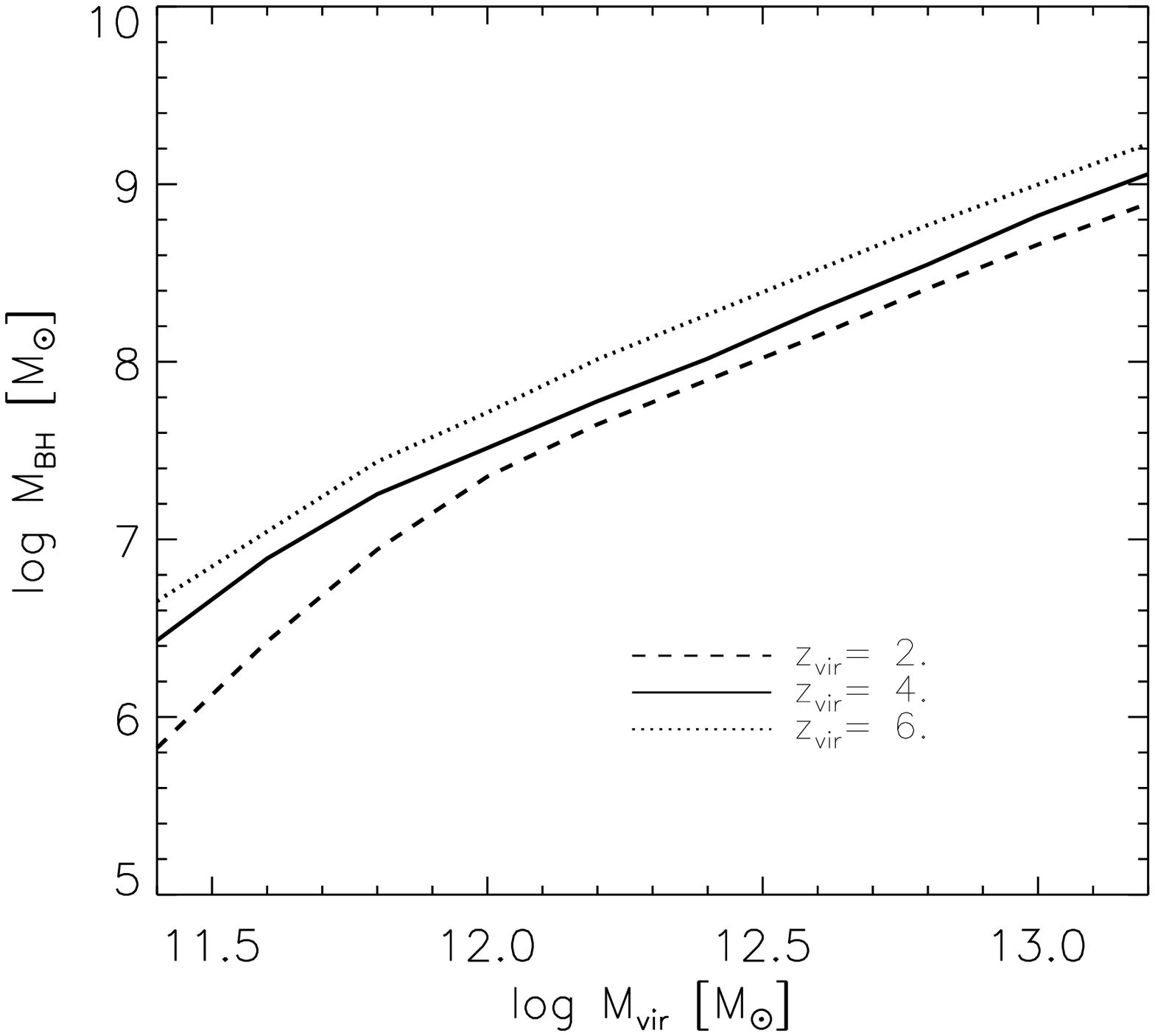}{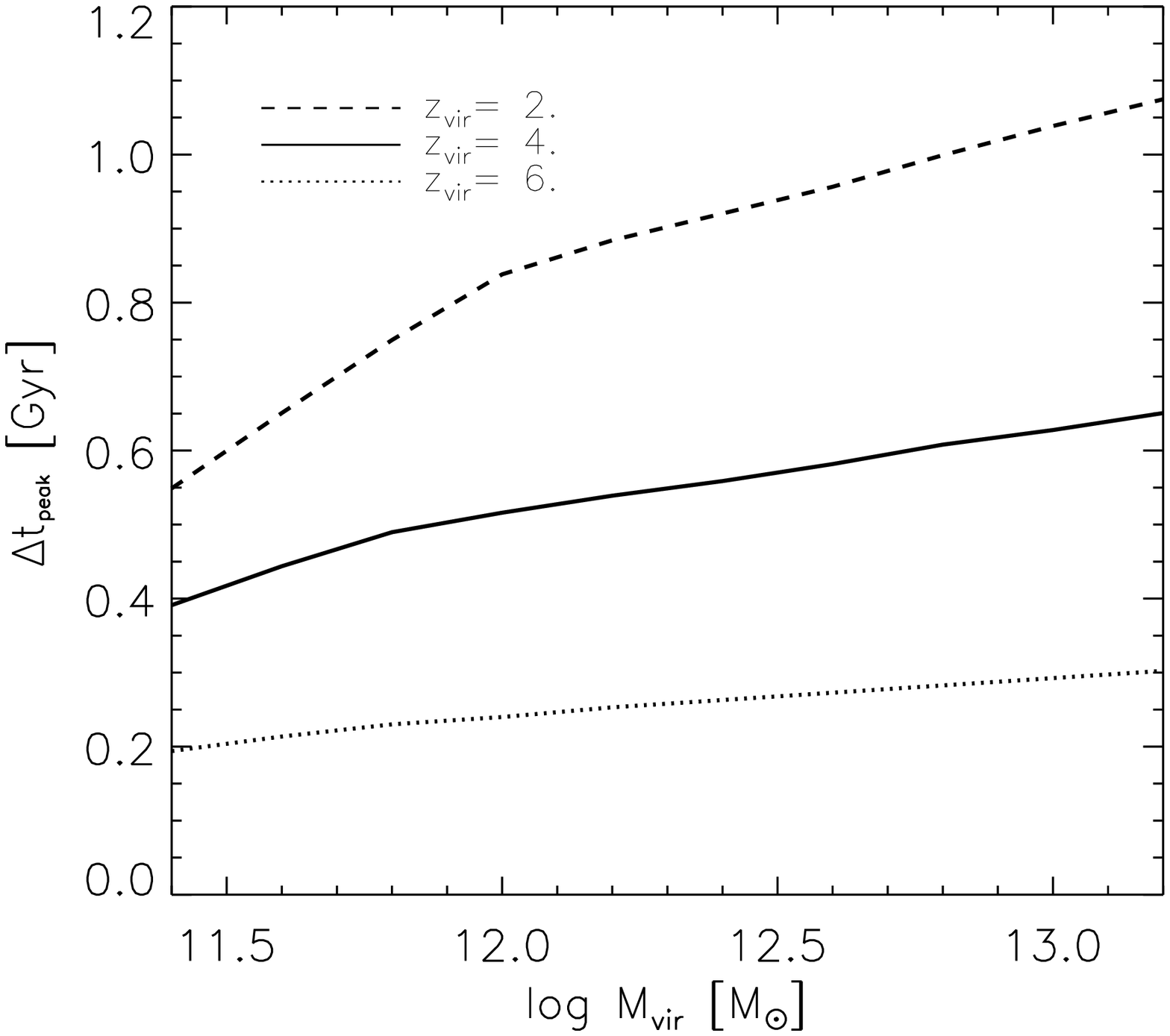}\caption{Left
panel: BH mass at the accretion peak as a function of the host
galaxy virial mass. Right panel: time lag between the host galaxy
virialization and the peak in the BH accretion rate, as a function
of the host virial mass.}\label{fig|Mbh&tdelay}
\end{figure}

\section{The model at work}

As a first step we want to properly translate the DM formation rates
given by Eq.~(\ref{eq|DMrates}) into BH formation rates. This calls
for a relation between the halo mass $M_{\mathrm{vir}}$ and the BH
mass $M_{\bullet}$. In Fig.~\ref{fig|Mbh&tdelay} (left panel) we
plot the one obtained by solving the equations presented in Appendix
A, that can be well approximated by
\begin{equation}
M_{\bullet} \approx 8\times 10^6\, {(M_{\mathrm{vir}}/2.2\times
10^{11}\, M_{\odot})^{3.97}\over 1+(M_{\mathrm{vir}}/2.2\times
10^{11}\, M_{\odot})^{2.7}}\, \left({1+z_{\mathrm{vir}}\over
7}\right)~M_{\odot}~~.\label{eq|Mbh_Mh}
\end{equation}
This relationship is very close to that found, at $z=0$, by Shankar
et al. (2006) by comparing the local BH mass function with the
galaxy halo mass function. However, this is only an average
relationship and we do expect a significant scatter around it, since
the values of the parameters in the equations of Appendix A may
naturally vary from halo to halo. In the following we assume that
the above relation $\log{M_{\bullet}}-\log{M_{\mathrm{vir}}}$ holds
on average with a gaussian dispersion $\Delta \log M_{\bullet}$.
Therefore we convert the halo formation rates into BH formation
rates through the convolution
\begin{equation}
{\mathrm{d}^2\, N_{\mathrm{BH}}\over \mathrm{d} t_{\mathrm{vir}}\,
\mathrm{d} M_\bullet} = \int{\mathrm{d}\log{M_\bullet'}}~
\left|{\mathrm{d} M_{\mathrm{vir}}\over \mathrm{d}
M_{\bullet}}\right|_{M_{\bullet}'} ~~ {\mathrm{d}^2\,
N_{\mathrm{ST}}\over \mathrm{d} t_{\mathrm{vir}}\, \mathrm{d}
M_{\mathrm{vir}}}_{\big|M_{\mathrm{vir}}(M_{\bullet}')}\,
{e^{-(\log{M_\bullet'}-\log{M_\bullet})^2/2\, (\Delta
\log{M_\bullet})^2}\over
\sqrt{2\pi\,(\Delta\log{M_\bullet})^2}}~~.\label{eq|BHrates}
\end{equation}
It will be shown below that the dispersion is an important
ingredient when comparing model predictions to observations.

Note that despite the redshift dependence of the
$M_{\bullet}/M_{\mathrm{vir}}$ ratio (see Eq.~[\ref{eq|Mbh_Mh}]),
our model predicts a relatively weak evolution of the BH-bulge mass
ratio. This is because, as mentioned in \S~2.2, the growth of the BH
mass is controlled by the SFR through the radiation drag and the SN
feedback; in turn, the feedback from the QSO can eventually sweep
out the gas halting both the star formation and the accretion on the
BH, especially in massive galaxies. In the latter structures the
stellar and BH masses grow (and stop growing) in parallel, while in
smaller objects where only SN feedback is effective the BH-bulge
mass ratio turns out to depend slightly on $z$ (see Fig.~5 of
Cirasuolo et al. 2005).

\subsection{QSO luminosity functions and BH mass function}

The QSO LFs can now be computed. Up to its peak, the BH bolometric
light curve can be well approximated by the simple exponential form
\begin{equation}
L(t) = {\lambda\, M_{\bullet}\, c^2\over t_{\mathrm{Edd}}}\,
e^{(t-t_{\mathrm{vir}}-\Delta
t_{\mathrm{peak}})/\tau_{\mathrm{ef}}}\,
\theta_H\bigl(t_{\mathrm{vir}}+\Delta t_{\mathrm{peak}}-\Delta
t_{\mathrm{vis}}\la t\la t_{\mathrm{vir}}+\Delta
t_{\mathrm{peak}}\bigr)~~.\label{eq|lightcurve}
\end{equation}
Here $t_{\mathrm{Edd}}\approx 4\times 10^8$ yr is the Eddington
timescale, and $\tau_{\mathrm{ef}}\approx \eta\,
t_{\mathrm{Edd}}/(1-\eta)\, \lambda$ is the $e$-folding time in
terms of the BH mass-energy conversion efficiency $\eta$ and of the
Eddington ratio $\lambda$. The Heaviside function\footnote{The
Heaviside function $\theta_H$ is defined by
\[
\theta_H (x)= \left\{
  \begin{array}{ll}
    1, & \mathrm{if}~~ x~~ \mathrm{is~~true}; \\
    0, & \mathrm{otherwise}.
  \end{array}
\right.~.
\] }
$\theta_H$ specifies that the QSO shines unobscured only during the
time interval $\Delta t_{\mathrm{vis}}$ before the peak of its light
curve. This means that, as discussed in \S~2.2, at the time
$t_{\mathrm{vir}}+\Delta t_{\mathrm{peak}}-\Delta t_{\mathrm{vis}}$
the BH is massive enough for its energy feedback to remove most of
surrounding gas and dust, letting the active nucleus to be visible.
In this work we have used an Eddington ratio $\lambda$ slightly
rising toward high $z$, and specifically $\lambda=4$ for $z\ga 6$,
$\lambda=3$ for $5\la z\la 6$, $\lambda=1.7$ for $3\la z\la 5$,
$\lambda=1$ for $2\la z\la 3$, and $\lambda=0.8$ for $1.5\la z\la
2$. The empirical formula $\lambda(z)\approx -1.15+0.75\, (1+z)$
works as well in the redshift range $1.5\la z\la 6$ (see also
Appendix A and Table~A1). This assumption is discussed in \S~6.

There are three timescales in Eq.~(\ref{eq|lightcurve}): the virial
time $t_{\mathrm{vir}}$ depends on cosmology; the peak time $\Delta
t_{\mathrm{peak}}$  is obtained by solving the system of equations
reported in Appendix A; the visibility time $\Delta
t_{\mathrm{vis}}$, dependent on dust and gas absorption, is taken as
a parameter, since its computation is challenging for
semi-analytical models.

The QSO LF at a time $t$ and luminosity $L$ is computed by summing
up the contributions of all the sources which virialize at epochs
$t_{\mathrm{vir}} \la t$ and shine at the time $t$ with luminosity
$L$. One has
\begin{equation}
\Phi(L,t) = \int_{t- \Delta t_{\mathrm{peak}}}^{t- \Delta
t_{\mathrm{peak}}+\Delta
t_{\mathrm{vis}}}{\mathrm{d}t_{\mathrm{vir}}}\, \int{\mathrm{d}
M_\bullet~ {\mathrm{d}^2\, N_{\mathrm{BH}}\over \mathrm{d}
t_{\mathrm{vir}}\, \mathrm{d} M_\bullet}}~~
\delta_D\bigg(L-{\lambda\, M_{\bullet}\, c^2\over
t_{\mathrm{Edd}}}\, e^{(t-t_{\mathrm{vir}}-\Delta
t_{\mathrm{peak}})/\tau_{\mathrm{ef}}}\biggr)~~,\label{eq|QSOLF}
\end{equation}
where $\delta_D$ indicates the Dirac delta function. The time delay
$\Delta t_{\mathrm{peak}}$ provided by our code ranges from $0.2$
Gyr at redshift $z\ga 5$ where $\lambda\approx 4$ to $1$ Gyr at
redshifts $z\la 2$, where $\lambda\la 1$ (see
Fig.~\ref{fig|Mbh&tdelay}, right panel).

In our model, the declining phase of the QSO light curve can be
neglected for halo masses above few $\times 10^{12}\, M_{\odot}$,
because for them the QSO feedback is so powerful to remove suddenly
most of the accreting gas from the surroundings. As a consequence,
after the peak the accretion rate drops abruptly to values much
lower (by factors of $10^{-2}$ at least; see Fig.~\ref{fig|star&BH})
than the Eddington rate, and the radiation process becomes much less
efficient (for a review see Blandford 2004). We have checked that
such radiatively inefficient stages provide a negligible
contribution to the QSO LFs, in accord with the findings by Yu \& Lu
(2004) based on the comparison between the observed QSO LFs and the
local supermassive BH mass function.

On the contrary, for halo masses smaller than $10^{12}\, M_{\odot}$
and especially at redshifts $z\la 2$, the BH accretion rate decays
more gently after the peak, and up to few Gyrs are required before
it lowers to levels of $\sim 10^{-2}$ the Eddington rate, becoming
radiatively inefficient. This declining phase is consistent with
observations of moderate QSO activity in relatively typical
early-type galaxies, with old stellar populations (e.g., Kauffmann
et al. 2003). This regime, which adds to the visibility time defined
above (that refers to the rising phase of the bolometric light
curve), is included in our computation of the QSO LFs. We find that
it contributes significantly to the faint end of the X-ray QSO LFs
(see also Hopkins et al. 2006b) and to the X-ray number counts (see
Granato et al. 2006 for details). On the other hand, the optical
luminosities are generally below those corresponding to the standard
definition of optical QSOs ($M_B \le -23$); therefore the declining
phase has a lower effect on optical LFs.

The last step of our computation is the conversion of the bolometric
QSO LFs computed above to the optical and hard X-ray bands through
the appropriate bolometric corrections (e.g., Elvis et al. 1994;
Ueda et al. 2003; Vignali, Brandt \& Schneider 2003; McLure \&
Dunlop 2004; Barger et al. 2005; Richards et al. 2005, 2006). For
the optical band we use $L/L_B=f_B$ with $f_B\approx 10$. For the
hard X-ray band ($2$--$10$ keV) we take into account the observed
luminosity dependence of $L/L_X=f_X$, given by $f_X\approx
k_X^{1/(\beta_X+1)}\, (L/10^{43}\, \mathrm{erg~
s}^{-1})^{\beta_X/(\beta_X+1)}$, with $k_X\approx 17$ and
$\beta_X\approx 0.43$ (see also Shankar et al. 2004).

The epoch-dependent BH mass function is given by
\begin{equation}
\Psi(\log{M_{\bullet}},t) = \int_0^{t}{\mathrm{d}
t_{\mathrm{vir}}}~~{\mathrm{d}^2\, N_{\mathrm{BH}}\over \mathrm{d}
t_{\mathrm{vir}}\, \mathrm{d}
\log{M_\bullet}}~~.\label{eq|BHmassfun}
\end{equation}

\subsection{Related galactic properties}

Our model aims at predicting, at the same time, both the QSO LFs at
various redshifts, and the properties of host galaxies.

Using the DM halo formation rates of Eq.~(\ref{eq|DMrates}) we can
compute the galaxy LF in a given band as
\begin{equation}
\Phi(\log L_{\mathrm{i}},t) = \int_0^{t}{\mathrm{d}
t_{\mathrm{vir}}}~~{\mathrm{d}^2\, N_{\mathrm{ST}}\over \mathrm{d}
t_{\mathrm{vir}}\, \mathrm{d} M_{\mathrm{vir}}}\, \left|{\mathrm{d}
M_{\mathrm{vir}}\over \mathrm{d}\log
L_{\mathrm{i}}}\right|~~,\label{eq|GalLF}
\end{equation}
where $L_{\mathrm{i}}$ is the luminosity in the i-th band.
Similarly, the velocity distribution function is given by
\begin{equation}
\Omega(\log{\sigma},t) = \int_0^{t}{\mathrm{d}
t_{\mathrm{vir}}}~~{\mathrm{d}^2\, N_{\mathrm{ST}}\over \mathrm{d}
t_{\mathrm{vir}}\, \mathrm{d} M_{\mathrm{vir}}}\, \left|{\mathrm{d}
M_{\mathrm{vir}}\over \mathrm{d}
\log{\sigma}}\right|~~.\label{eq|GalVDF}
\end{equation}
To derive the Jacobians we use: (i) the GRASIL code, which evaluates
the luminosity of a galaxy as a function of time once the dust
content and the SFR (obtained by solving the system of equations in
Appendix A) is specified; (ii) the scaling law for the virial
velocity $V_{\mathrm{vir}}\equiv
(G\,M_{\mathrm{vir}}/R_{\mathrm{vir}})^{1/2}\propto
M_{\mathrm{vir}}^{1/3}\,(1+z_{\mathrm{vir}})^{1/2}$; (iii) the ratio
between virial velocity and stellar velocity dispersion,
$\sigma=0.55\,V_{\mathrm{vir}}$, found by Cirasuolo et al. (2005).
The combination of these three ingredients yields a
magnitude-velocity dispersion (Faber-Jackson) relation in good
agreement with that ($M_{r^*} = -7.5\, \log{\sigma}-5$)
observationally derived by Sheth et al. (2003) through the bisector
method (see also Cirasuolo et al. 2005).

Exploiting the $\Phi(L_{\mathrm{i}},t)$, it is straightforward to
compute the galaxy counts and the related redshift distributions.
These estimates have been worked out in particular for sub-mm and
K-band selected galaxies, whose counts and redshift distributions
proved to be particularly challenging for competing semi-analytic
models. We defer the interested reader to the papers by Granato et
al. (2004) and Silva et al. (2005) for the description of the
computational strategy, based on the GRASIL code, and of the
results.

\begin{figure}[t]
\epsscale{1.}\plottwo{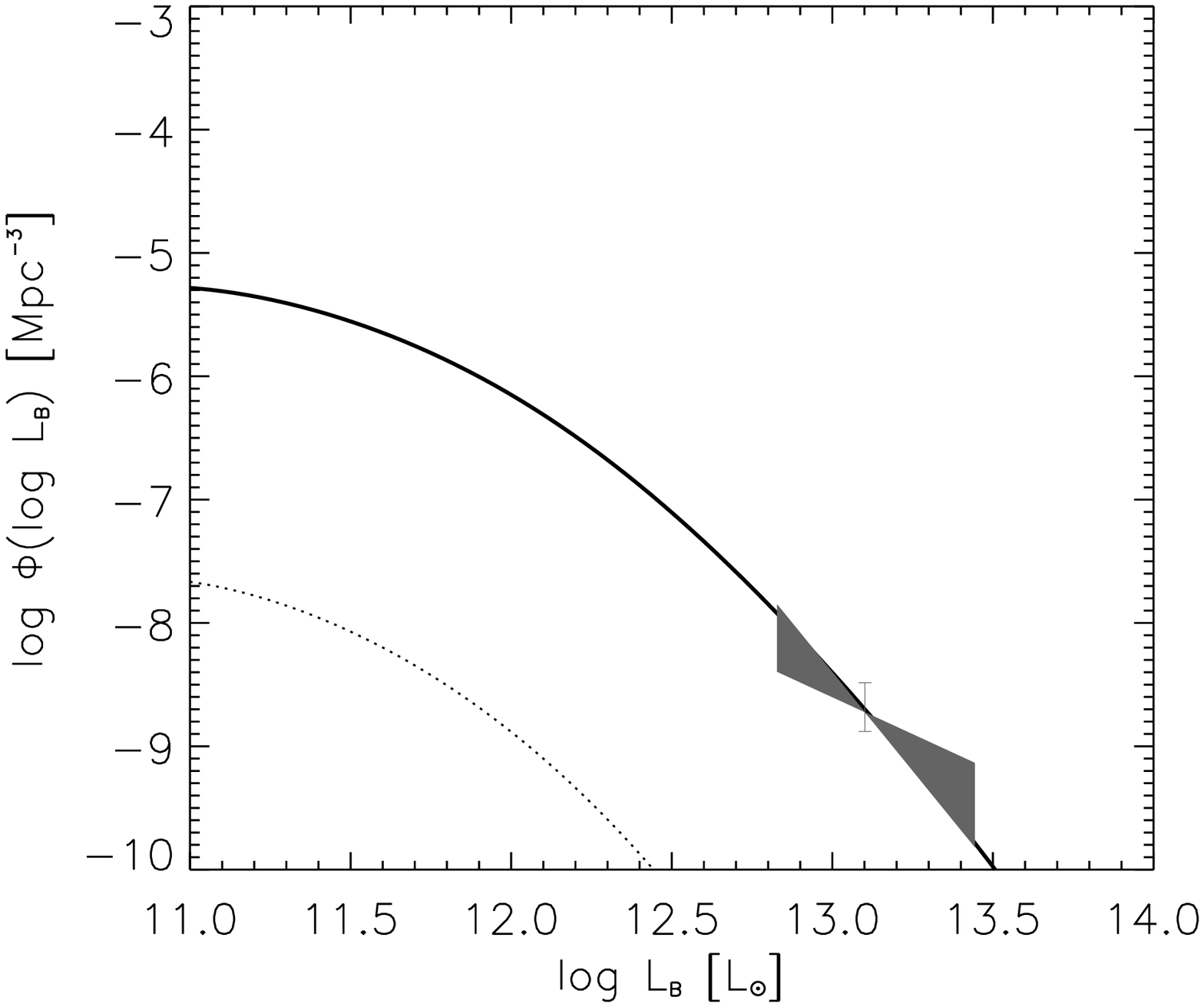}{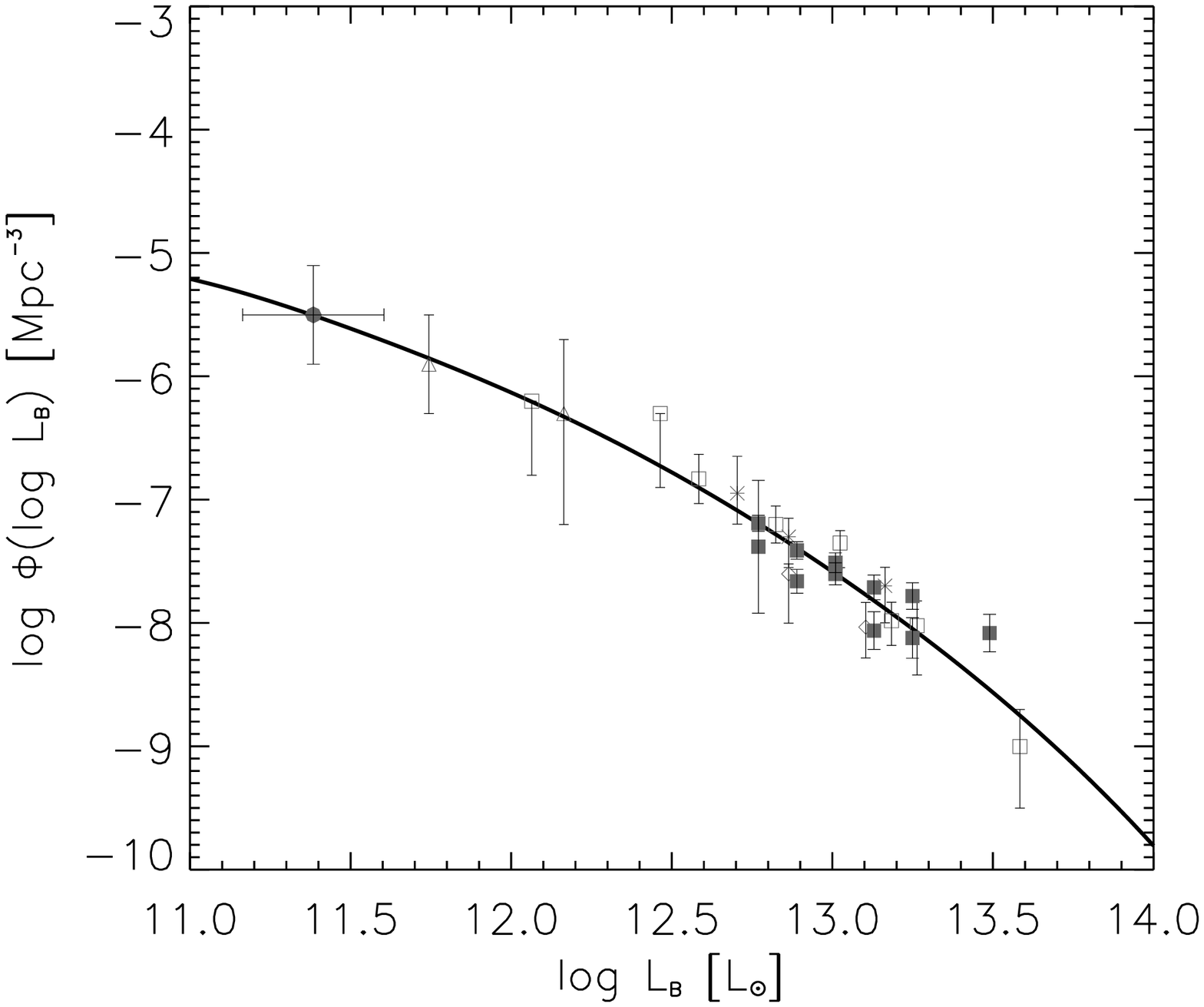}\caption{Left
panel: predicted QSO LF in the optical B-band at $z=6$
(\textit{solid} line) and $z=8$ (\textit{dotted} line) for an
Eddington ratio $\lambda=4$, and a visibility time $\Delta
t_{\mathrm{vis}}\approx 5\times 10^7$ yr. Data points are from Fan
et al. (2004). Right panel: QSO LF in the optical B-band at $z=5$
for $\lambda=3$ and $\Delta t_{\mathrm{vis}}\approx 5\times 10^7$
yr. Data points are from Kennefick et al. (1995, \textit{open
squares}); Fan et al. (2001, \textit{diamonds} and
\textit{asterisks}); Wolf et al. (2003, \textit{triangles});
Cristiani et al. (2004, \textit{circles}); Richards et al. (2006,
\textit{filled squares}).}\label{fig|QSO_highz}
\end{figure}

\section{Results}

\subsection{QSO Luminosity Functions}

An important feature of the model is the prediction of a pre-QSO
phase lasting $\Delta t_{\mathrm{peak}}\sim 15-20\, e$-folding times
(for the most massive objects), during which the mass of the central
BH is increasing exponentially. Meanwhile the host galaxy is forming
stars at extremely high rates (cf. Appendix A and
Fig.~\ref{fig|star&BH}; see also Sanders et al. 1988; Croom et al.
2006; Hopkins et al. 2006) and is therefore very bright at far-IR
and (sub-)mm wavelengths.

\begin{figure}[t]
\epsscale{1.}\plottwo{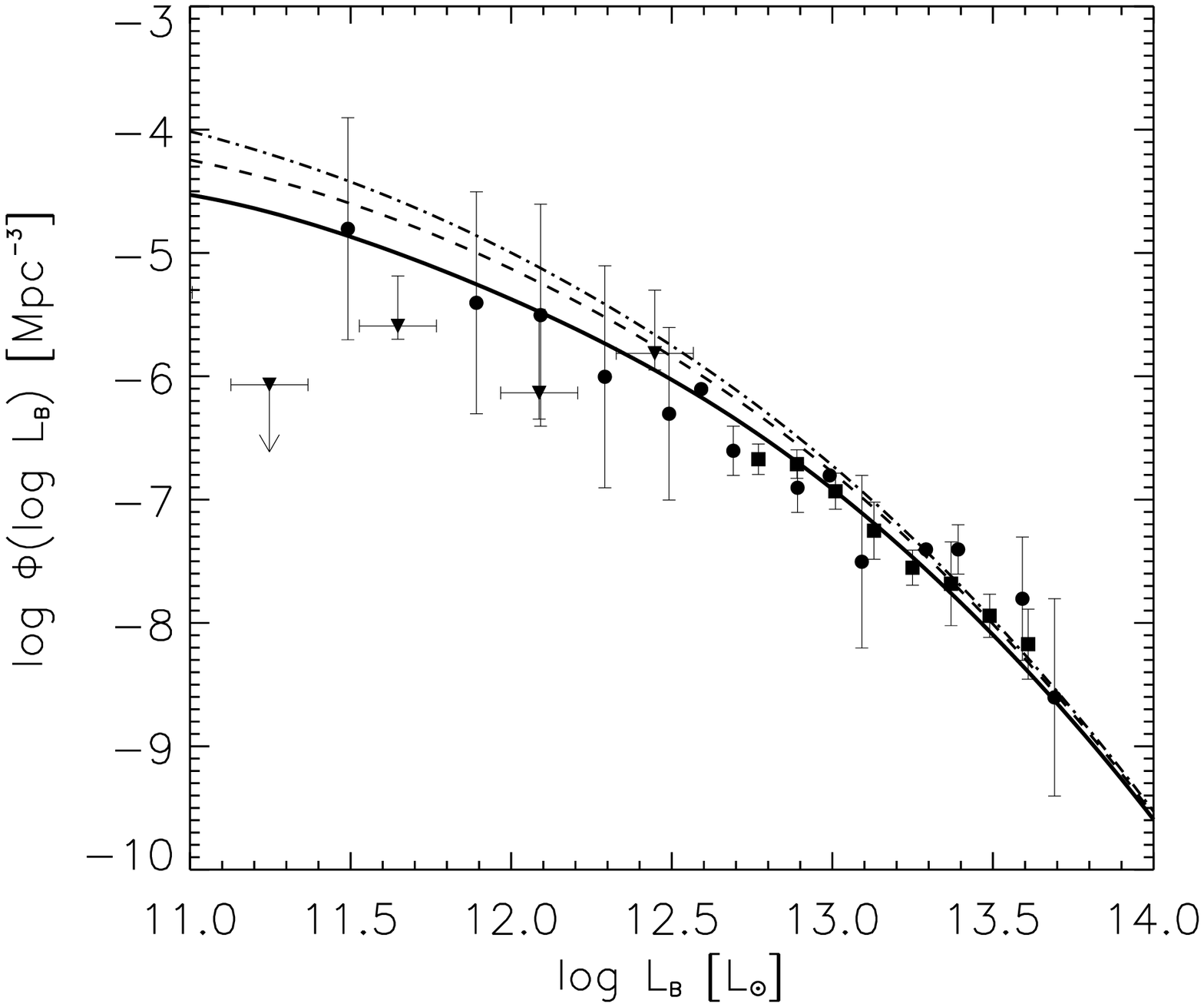}{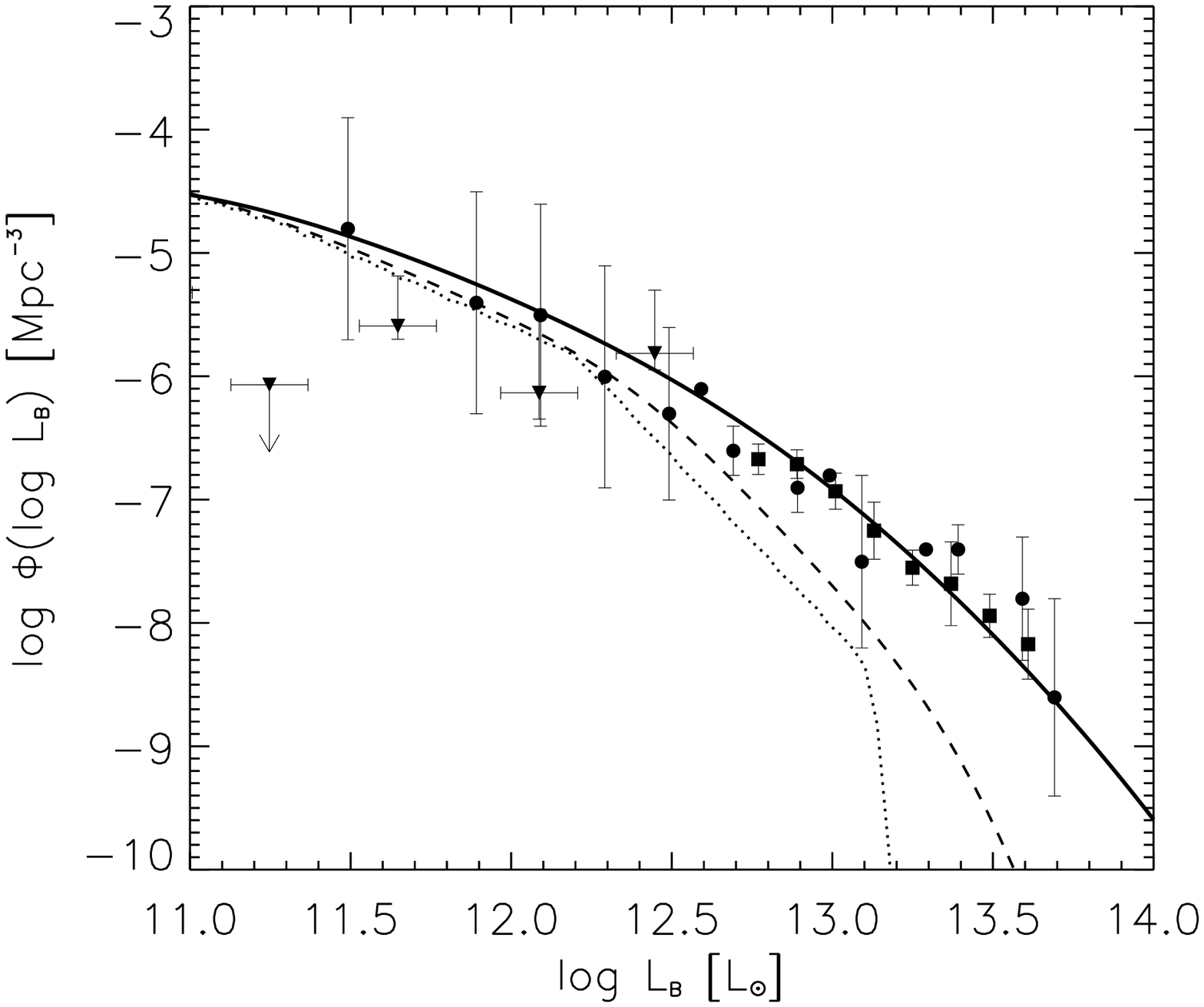}\caption{Left
panel: predicted QSO LF in the optical B-band at redshift $3$ for
$\lambda=1.7$ and different values of the visibility time: $\Delta
t_{\mathrm{vis}}\approx 5\times 10^7$ yr (best fit, solid line),
$\Delta t_{\mathrm{vis}}\approx 10^8$ yr (dashed) and $\Delta
t_{\mathrm{vis}}\approx 3\times 10^8$ yr (dot-dashed). Right panel:
effect of the scatter around the mean
$\log{M_{\bullet}}-\log{M_{\mathrm{vir}}}$ relation. The solid curve
is the same as in the left panel and corresponds to a scatter of 0.3
dex, as assumed throughout this paper. The dashed curve is for a
scatter of 0.15 dex and the dotted one for zero scatter. In both
panels data are from Pei (1995, \textit{circles}), Hunt et al.
(2004, \textit{triangles}), and Richards et al. (2006,
\textit{squares}).} \label{fig|QSO_LF_himedz}
\end{figure}

\begin{figure}[t]
\epsscale{1.}\plottwo{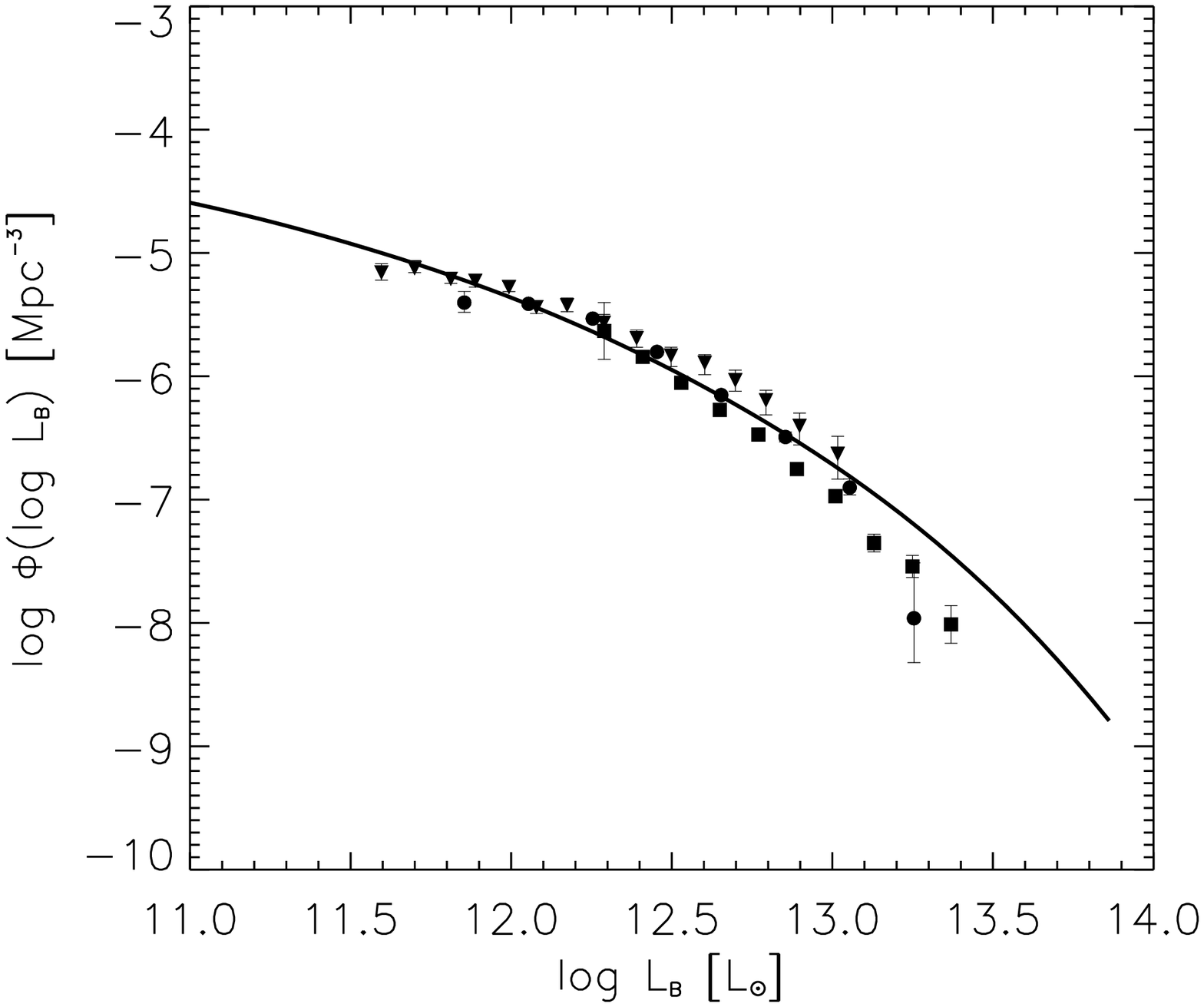}{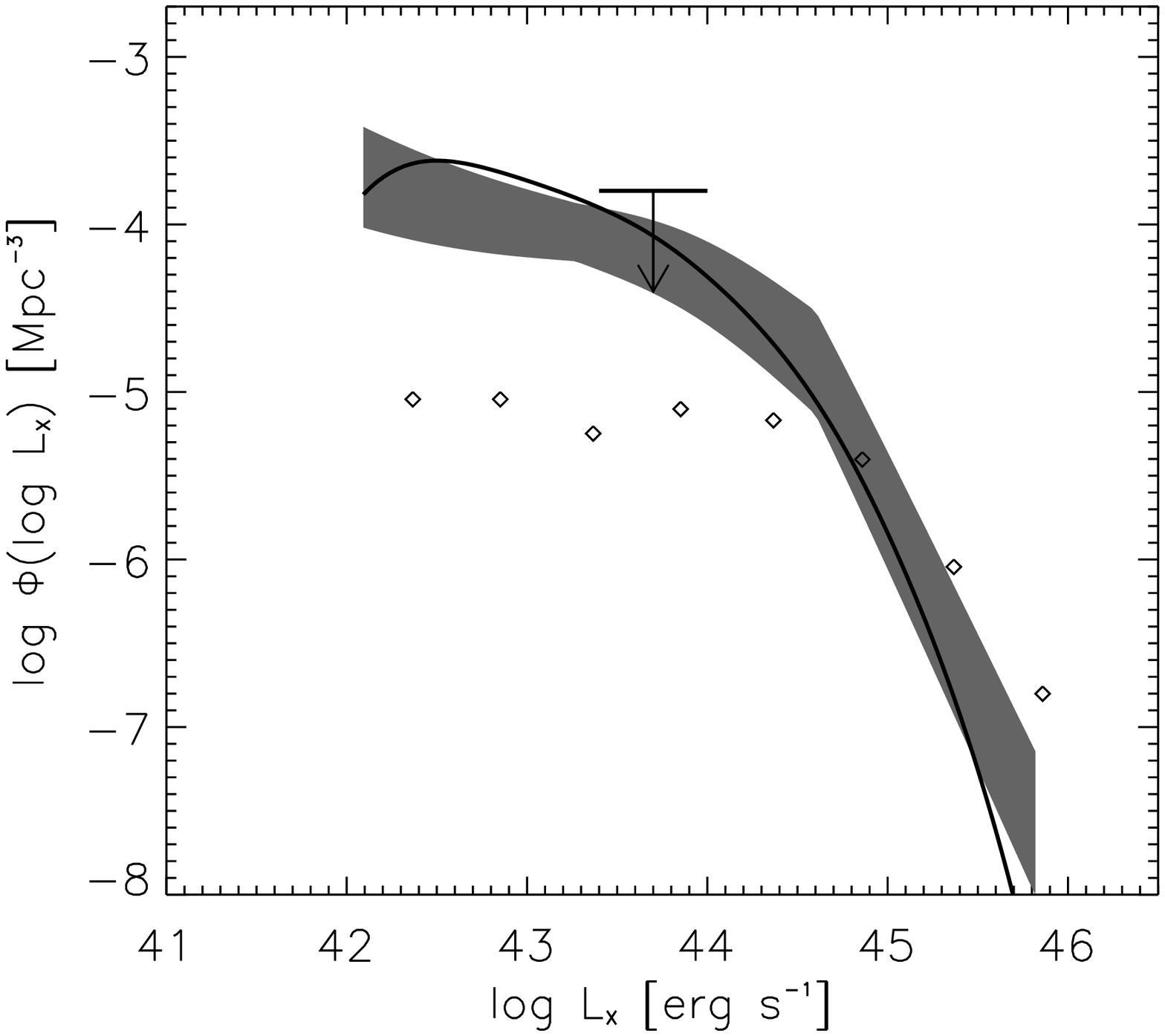}\caption{Left
panel: The QSO LF in the optical B-band at redshift $2$ where
$\lambda=1$, with the visibility time set to $\Delta
t_{\mathrm{vis}}\approx 10^8$ yr; data points are from Croom et al.
(2004; \textit{circles}), from Richards et al. (2005;
\textit{triangles}), and from Richards et al. (2006;
\textit{squares}). Right panel: The QSO LF in the hard X-ray band
(2-10 keVs) at redshift $2$ where $\lambda=1$, with the visibility
time set to $\Delta t_{\mathrm{vis}}\approx 3\times 10^8$ yr; data
are from Ueda et al. (2003, \textit{shaded area}), from Barger et
al. (2005, \textit{diamonds}), and from La Franca et al. (2005,
\textit{arrow}).} \label{fig|QSO_lowmedz}
\end{figure}

\begin{figure}[t]
\epsscale{1.}\plottwo{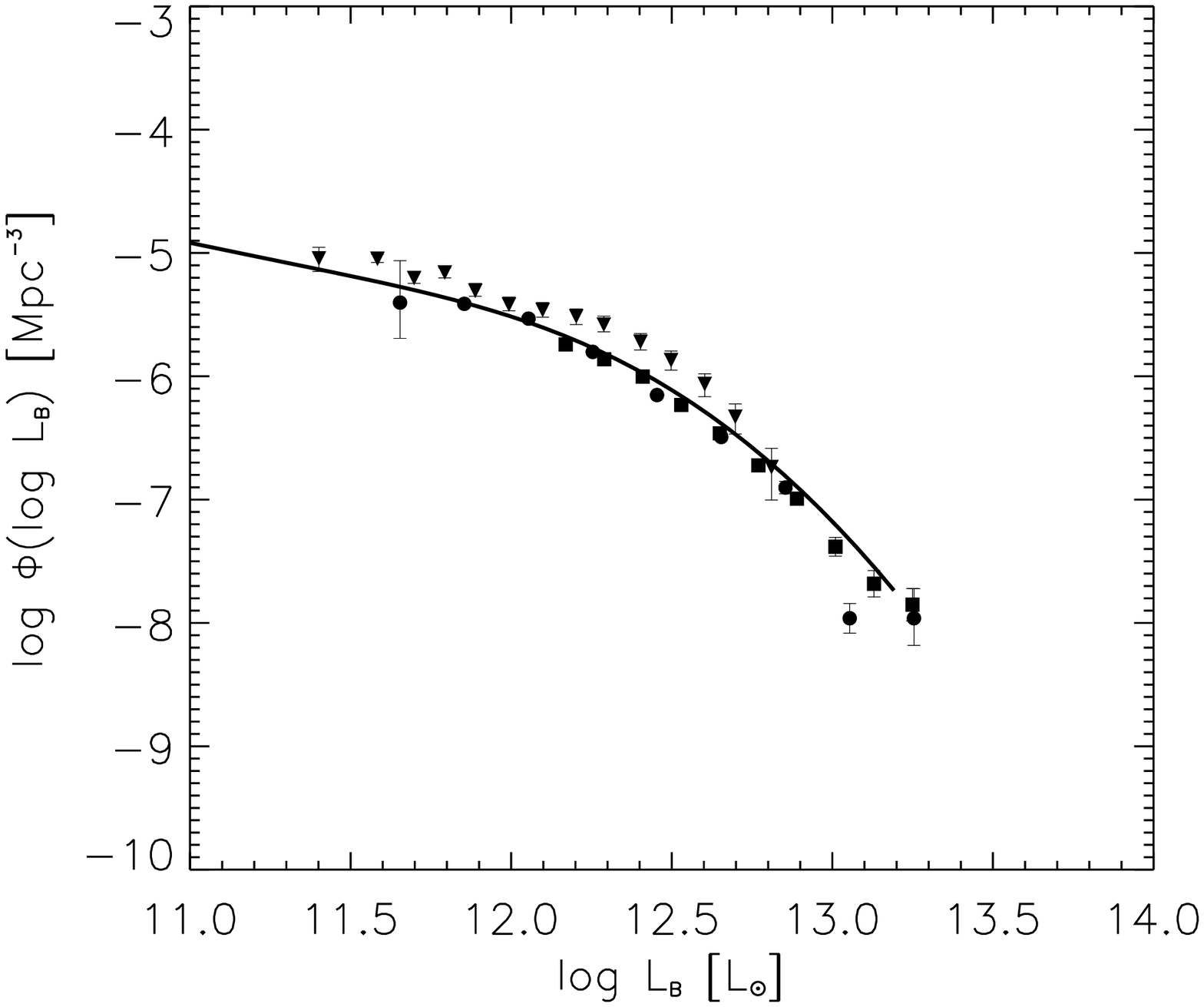}{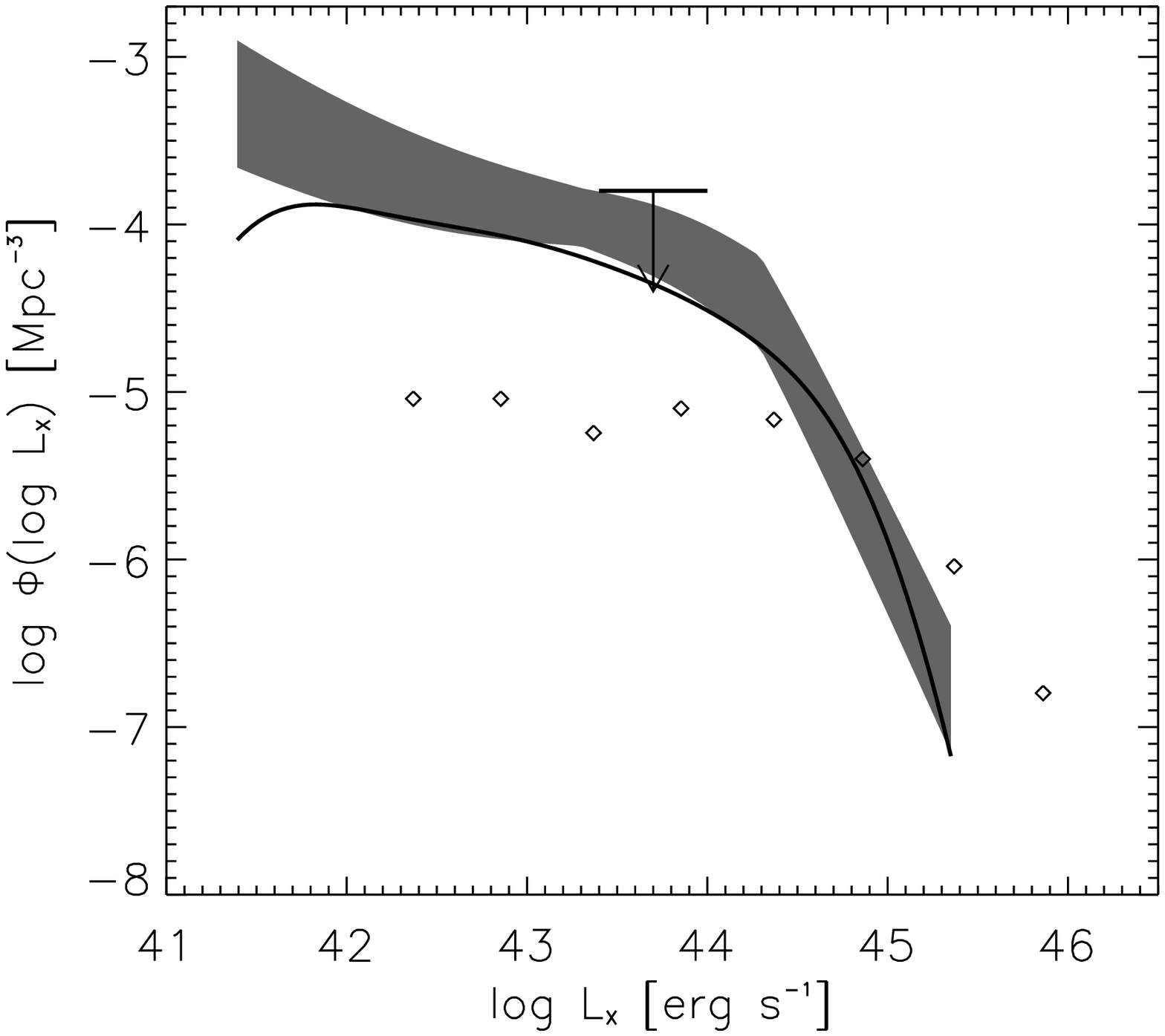}\caption{Left
panel: QSO LF in the optical B-band at $z=1.5$, for $\lambda=0.8$
and $\Delta t_{\mathrm{vis}}\approx 10^8$ yr. The data are from
Croom et al. (2004; \textit{circles}), Richards et al. (2005;
\textit{triangles}), and Richards et al. (2006; \textit{squares}).
Right panel: QSO LF in the hard X-ray band (2-10 keV) at $z=1.5$ for
$\lambda=0.8$ and $\Delta t_{\mathrm{vis}}\approx 3\times 10^8$ yr.
The data are from Ueda et al. (2003, \textit{shaded area}), Barger
et al. (2005, \textit{diamonds}), and La Franca et al. (2005,
\textit{arrow}).}\label{fig|QSO_lowz}
\end{figure}

In Figs.~\ref{fig|QSO_highz}-\ref{fig|QSO_lowz} we show that our
model provides a very good fit to the optical and hard X-ray QSO LFs
over the full redshift range over which it applies ($z \ga 1.5$).
Due to the delay $\Delta t_{\mathrm{peak}}$ between the halo
virialization epoch and the peak of the BH accretion, the QSOs
shining at the cosmic time $t_H(z)$ are associated to halos
virializing at an earlier time $t_{\mathrm{vir}}\approx
t_H(z)-\Delta t_{\mathrm{peak}}<t_H(z)$.

This is an important point since, at high redshifts and for massive
halos, $\Delta t_{\mathrm{peak}}$ corresponds to a substantial
fraction of the cosmic time (see Fig.~\ref{fig|Mbh&tdelay}, right
panel), and the DM halo formation rates at $t_{\mathrm{vir}}$ are
significantly lower than at $t_H(z)$. This is illustrated by
Fig.~\ref{fig|QSO_highz} (left panel), where we also show the
predicted optical LF of QSOs at $z=8$. The precipitous drop of the
number density of bright QSOs, compared to $z=6$, is due to the fact
that, for $t_H(z=8)$, $\Delta t_{\mathrm{peak}}\approx 0.2\,$Gyr
corresponds to $\Delta z \approx 2$, so that the abundance of bright
QSOs reflects that of virialized massive halos at $z\approx 10$,
which is far lower than at $z\approx 8$.

The flattening of the LFs at the low end is due to the flatter slope
of the formation rate of less massive halos (see
Eq.~[\ref{eq|DMrates}]) and to the weak dependence of the peak time,
$\Delta t_{\mathrm{peak}}$, on the halo mass (cf.
Fig.~\ref{fig|Mbh&tdelay}).

The highest luminosity portion of the LFs corresponds to the high
tail of the distribution of $M_{\bullet}$ around the mean
$\log{M_{\bullet}}-\log{M_{\mathrm{vir}}}$ relation, at large halo
masses. The fit to the data is obtained with a Gaussian distribution
with dispersion $\Delta \log{M_{\bullet}}\approx 0.3$.  As can be
seen in Fig.~\ref{fig|QSO_LF_himedz} (right panel), for $\Delta
\log{M_{\bullet}}\approx 0$ the cutoff in galaxy halo mass (cf.
\S~2.1) would yield a drastic drop around $L_B\approx
10^{13.2}\,L_{\odot}$. We have checked that, for a given mass and
virialization redshift of the host halo, a reasonable variation of
the physical parameters from structure to structure can account for
such a scatter. For example, at $z\approx 6$ and
$M_{\mathrm{vir}}\approx 2\times 10^{13}\, M_{\odot}$ it can be
achieved if the clumping factor $\mathcal{C}$ varies by a factor of
2, or the seed BH mass varies by a factor of 10, or the strength of
the QSO feedback $\epsilon_{QSO}$ varies by a factor of a few.
Mahmood et al. (2005) also noted this problem, and empirically
solved it by inserting a Lorentzian tail in the BH formation rate
for halo masses above $10^{13.2}\, M_{\odot}$.

The observations by Richards et al. (2006, see their Fig.~20) show
that the number density of very luminous QSOs with $M_{1450}\la -27$
peaks between $z=2$ and $z=3$. Our model reproduces this trend.
Specifically, the rise from high redshift to $z\approx 2.5$ is due
to the strong increase of the formation rate of very massive halos,
which overwhelms the effect of the decrease of the BH mass
associated to a given $M_{\mathrm{vir}}$ (see
Eq.~[\ref{eq|Mbh_Mh}]). But at redshifts $z\la 2$ the latter effect
dominates, causing the fall in the number density of bright sources.

To reproduce the optical data, we adopt a visibility time $\Delta
t_{\mathrm{vis}} \approx 5\times 10^7-10^8$ yr, the longer value
being favored at redshifts $z\la 2$. The X-ray LFs require
visibility times around $\Delta t_{\mathrm{vis}} \approx 3\times
10^8$ yr at $1.5\la z\la 3$. The visibility time turns out to be a
factor of $5-10$ shorter than the time $\Delta t_{\mathrm{peak}}$
spent by the BH to grow to its final mass. This implies that
absorption has a key role in determining the duration of the visible
QSO phase; in fact, large amounts of dust are present in the pre-QSO
phase, as witnessed by sub-mm observations. Note that such a phase
can also be probed through hard X-ray observations, than can detect
the emission from the growing supermassive BH. Actually such
emission has already been revealed from the center of sub-mm
selected galaxies (Alexander et al. 2005; Borys et al. 2005); a
specific discussion of this issue is presented by Granato et al.
(2006).

As stated in \S~3.1, the accretion rate onto the central BH has been
taken to be at most slightly super-Eddington, $\lambda =
\dot{M}_{\bullet}/\dot {M}_{\bullet}^{\mathrm{Edd}} \la 4$, as
expected when it is limited by radiation pressure (see Small \&
Blandford 1992; King \& Pounds 2003). Observational evaluations of
the Eddington ratio $\lambda$ for high luminosity QSOs with $L\ga
10^{47}$ erg s$^{-1}$ at redshifts $z\ga 2$ have been obtained by
Warner, Hamann, \& Dietrich (2004) and Vestergaard (2004). These
authors find a significant fraction of QSOs radiating at mildly
super-Eddington powers. On the other hand, at lower redshifts and
luminosities the emission seems to be limited to $\lambda \la 1$
(McLure \& Dunlop 2004; Kollmeier et al. 2006). Note that these
results are to be taken cautiously. In fact, they are based on the
observed FWHM of the H$_{\beta}$, MgII and CIV broad lines and on an
empirical relation linking the size of the broad-line region to the
QSO luminosity. On top of that, a normalization to the
$M_{\bullet}-\sigma$ relation is needed to set the scaling factor in
the equation yielding the BH mass. This coefficient depends
crucially on the kinematic and geometrical properties of the broad
line region itself (see Onken et al. 2004). Thus the inferred
Eddington ratios suffer from large observational and systematic
uncertainties, which amount to a factor of up to 10 on individual
objects, and up to 4 on statistical samples (see Vestergaard \&
Peterson 2006). In our model a variation of the Eddington ratio
$\lambda$ with the redshift is necessary to fit the QSOs LFs (see
\S$\,6$ for a discussion of this point).

If radiation pressure keeps the BH growth at around the Eddington
limit, outflows with mass rates $\dot{M}_w \sim
\dot{M}_{\mathrm{Edd}}$ are expected, as shown in Appendix A. In
massive galaxies the outflows remove most of the cold and infalling
gas. The model predicts the existence of clouds of chemically
enriched gas flowing out from the host galaxies (the expelled cold
gas). Such outflows have indeed been detected by spectroscopic
studies of narrow absorption lines associated to QSOs (Srianand \&
Petitjean 2000; D' Odorico et al. 2004). The average metallicity of
the gas expelled from large galaxies ($M_{\mathrm{vir}} \ga
10^{12}\, M_{\odot}$) is $Z\sim 1-2\, Z_{\odot}$, a value that after
proper dilution contributes to the metal abundance $Z\sim
Z_{\odot}/3$ of the intergalactic medium in the central regions of
clusters (see the review by Voit 2004). Note, however, that gas
clouds coming from the innermost regions can have strongly
supersolar metallicities.

\begin{figure}[t]
\epsscale{.8}\plotone{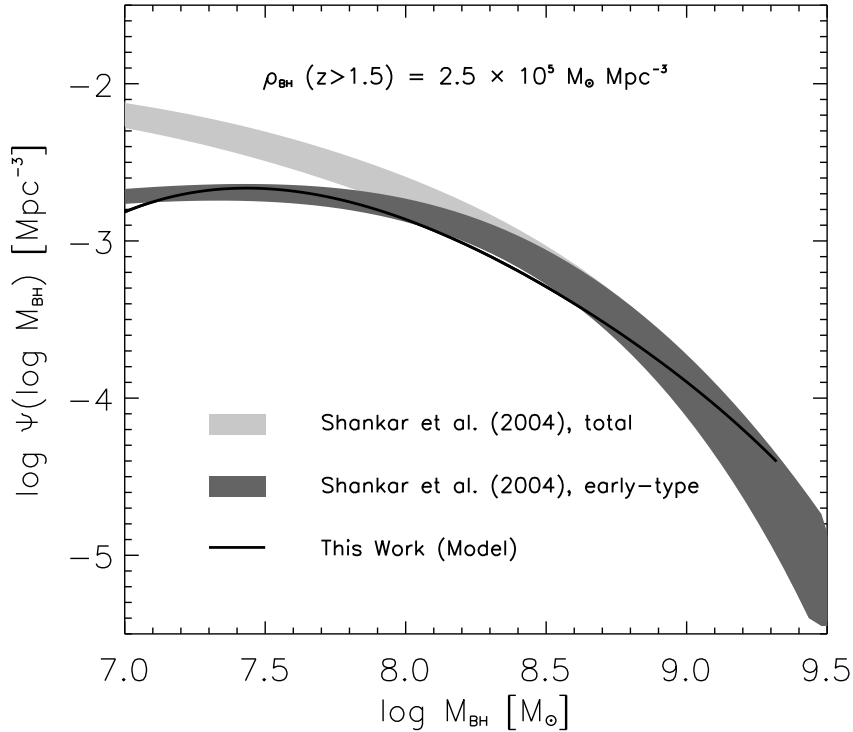}\caption{Comparison of the
predicted mass function of supermassive BHs associated with massive
spheroidal galaxies (solid line) with observational estimates
(Shankar et al. 2004). The dark shaded area refers to the local
supermassive BH mass function considering only early-type galaxies;
the light shaded area includes also bulges of late-types.}
\label{fig|QSO_demo}
\end{figure}

The accreted mass function predicted by our model is plotted in
Fig.~\ref{fig|QSO_demo}, and is found to be in good agreement with
the empirical determinations by Shankar et al. (2004) and Marconi et
al. (2004). This shows that about $60\%$ of the relic supermassive
BH mass density is in place already at $z\ga 1.5$. The rest of the
mass is accreted at later times onto BHs with final mass
$M_{\bullet}\la 10^8\, M_{\odot}$. This result agrees with
observations, showing that at $0.5 \la z\la 1.5$ QSOs of relatively
low luminosity, $L_X \la 3\times 10^{44}$ erg s$^{-1}$, provide
about $60\%$ of the X-ray background (Ueda et al. 2003). The $40\%$
in mass accreted at $z\la 1.5$ can account for this fraction if the
ratio of X-ray to bolometric luminosity increases with decreasing
redshift and/or luminosity, as suggested by observations (e.g.,
Wilkes 1994; Vignali et al. 2003; Strateva et al. 2005). The
accreted mass function is consistent with the Soltan (1982)
argument. We have checked this by redoing the calculations of
Shankar et al. (2004) with the Eddington ratios adopted in this
paper, and exploiting the Ueda et al. (2003) LFs corrected for very
obscured sources.

\subsection{Host galaxy properties}

The model, interfaced to the GRASIL code, provides also good fits to
a variety of galactic properties. Most of the results have been
presented in previous papers (Granato et al. 2004, Cirasuolo et al.
2005, Silva et al. 2005) and are summarized in Table~A2. Note that
the current set of parameters differs from the one adopted in
Granato et al. (2004), but the general scenario and in particular
the results concerning the galaxy properties are essentially
unaffected.

Figure~\ref{fig|star&BH} shows that small and large halos exhibit
quite different behaviors. In halos with masses $M_{\mathrm{vir}}\la
10^{12}\, M_{\odot}$ the interplay between SFR, BH accretion, and
respective feedback processes, limits the growth of the central BH,
which never reaches the power needed to remove all the gas. Thus the
SFR is prolonged over times $\Delta t_{\star} \ga \Delta
t_{\mathrm{peak}}$, i.e., a substantial amount of stars is formed
after the peak of the QSO activity. Contrariwise, in large galaxies
with $M_{\mathrm{vir}}\ga 10^{12} M_{\odot}$ the growth of the BH
produces winds, which are able to stop star formation in a short
time $\Delta t_{\star}\approx \Delta t_{\mathrm{peak}}$, see
Fig.~\ref{fig|Mbh&tdelay} (right panel).

On the other hand, in the pre-QSO phase the SFR varies from several
hundred to several thousand solar masses per year, within halos of
mass ranging from $10^{12}$ to $10^{13}\, M_{\odot}$. Almost all the
associated power is emitted in the far-IR. The model fits the $850\,
\mu$m counts; in particular, it predicts a large surface density
$N(\ga 1 \mathrm{mJy}, z\ga 5)\approx 600$ (sq. deg.)$^{-1}$ of
pre-QSO host galaxies at substantial redshift among bright objects
selected at $850 \, \mu$m (Silva et al. 2005). This large density
(compared to that of the QSOs at the same redshift) is mainly
ascribed to the fact that the pre-QSO phase lasts a factor of $10$
longer than the QSO phase, i.e., $\Delta t_{\mathrm{peak}}/\Delta
t_{\mathrm{vis}}\approx 10$ at high redshifts.

A direct consequence of the extremely high SFR in massive halos is
the early appearance of galaxies with large mass in stars
($M_{\star}\propto \mathrm{SFR} \times \Delta t_{\mathrm{peak}}$ at
high redshifts). These galaxies, at redshifts corresponding to
cosmological times $t_H\ga t_{\mathrm{vir}}+\Delta
t_{\mathrm{peak}}$, have already gone through the QSO phase and are
evolving passively, i.e., without major additions of newly formed
stars or large mass accretions onto the central BH.

Note that the $850\, \mu$m counts can be reproduced also adopting
SFRs much lower than predicted by our model, provided that a flatter
initial mass function is assumed (Baugh et al. 2005). The latter
approach yields a much lower number density of massive,
high-redshift galaxies. For example, the predicted number density of
galaxies with $M_{\star}\ga 10^{11}\, M_{\odot}$ becomes negligibly
small at redshift $z\ga 2.5$, while in our model it decreases by
only a factor of $2$ compared to that at $z\approx 1.5$, and it is
still significant at $z\approx 5$. Stronger constraints on this
number density would therefore help in discriminating between the
two scenarios (e.g., Tecza et al. 2004; Fontana et al. 2004; Greve
et al. 2005; Caputi et al. 2006; Bouwens \& Illingworth 2006).

Our model also naturally accounts for the observed trend of more
massive spheroids being richer in metals (see Thomas et al. 2005).
In our model the behavior of the average stellar metallicity
$\langle Z_*\rangle$ as a function of the host halo mass is well
described by the approximate expression $\langle Z_*\rangle \approx
-0.247+0.023\,\log{(M_{\mathrm{vir}}/M_{\odot})}$. This dependence
can be understood as follows. In massive galaxies the SN feedback
can remove only a small fraction of the cold gas involved in star
formation, which is rapidly metal-enriched due to the high SFR. When
the QSO reaches the peak of its activity, the energy released
removes from the halo most of both the enriched cold gas and of the
infalling metal-poor gas (see also Cox et al. 2006). Afterwards, the
gas infall and the star formation cease, and the metallicity of the
gas does not change anymore. In small galaxies SN explosions are
more efficient at removing the enriched cold gas but not the
infalling metal-poor gas. Thus the former can continuously be
replaced by the latter, and the metallicity of the cold gas remains
low. As a result large galaxies exhibit stellar populations with
larger metal content than smaller galaxies do. We also stress that
the short $\Delta t_{\star}\approx 0.2-0.5$ Gyr for
$z_{\mathrm{vir}} \ge 4$ is responsible for the observed enhancement
of $\alpha$ elements in massive galaxies (see also Romano et al.
2002; Granato et al. 2004), since the Fe enrichment is strictly
related to the explosions of Type-I$a$ SNae and these are delayed by
about $1$ Gyr.

The power of the QSO outflows ensures that only about $30\%$ of the
initial cosmic baryons are turned into stars in large halos
($M_{\mathrm{vir}}\ga 10^{12}\, M_{\odot}$). For smaller halos the
long lasting action of SNae is able to remove an even larger
fraction of gas, letting only less than $10\%$ of the initial gas to
turn into stars (Granato et al. 2004).

Cirasuolo et al. (2005) found that the observed velocity dispersion
function of spheroidal galaxies (Sheth et al. 2003) and the
Faber-Jackson relation can be reproduced under the hypotheses that
the old stellar populations are located in galaxy halos virialized
at redshift $z \ga 1.5$ and that the relation $\sigma \approx 0.55
\, V_{\mathrm{vir}}$ holds; here the velocity dispersion  $\sigma$
refers to the old stellar populations in the central galaxy regions.
They also showed (cf. their Fig.~4) that the same $\sigma
-V_{\mathrm{vir}}$ relation combined with the
$M_{\bullet}-M_{\mathrm{vir}}$ relation for redshift $z\ga 1.5$,
fits the data in the $M_{\bullet}-\sigma$ plane, with their
dispersion (Ferrarese \& Merritt 2000; Gebhardt et al. 2000;
Tremaine et al. 2002).

Our model predicts a steepening of the relationship in this plane at
low masses, since for small halos SNae damp down the SFR, thus
hampering the growth of the reservoir around the seed BHs and
eventually of the BHs themselves (cf. Eq.~[\ref{eq|snfeed}]; see
also Granato et al. 2004). This steepening is fully consistent with
the upper limits on the BH mass for M33 (Gebhardt et al. 2001) and
NGC205 (Valluri et al. 2005) obtained by direct dynamical
measurements, although the indirect determinations by Greene \& Ho
(2006) suggest a constant slope and a wide scatter at small BH
masses. On the other hand, we should also keep in mind that the low
BH mass portion of the diagram might just reflect the distribution
of BH seeds possibly created by merging of smaller BHs during the
fast accretion epoch and be only weakly affected by the mass
accretion. In addition, low BH masses may be easily increased by
substantial factors during later accretion phases, when host
galaxies are disk-(rather than bulge-)dominated.

\section{Comparison with previous models}

Accounting for the wealth of data on the QSO and host galaxy LFs and
on their statistical properties and relationships is a severe
challenge for analytical, semi-analytical and numerical galaxy
formation and evolution models.

The standard methodology adopted by analytical models to estimate
the QSO LFs is to convert the DM halo formation rates into BH
formation rates through a relationship between the mass of the
supermassive BH and that of the host halo. Then a description of the
QSO light curve (usually assuming Eddington-limited accretion) is
adopted to derive the QSO luminosity. The QSO LFs are then built up
according to the approximate expression
\begin{equation}
\Phi(L,z)\approx \Delta t_{\mathrm{vis}}\, {\mathrm{d}^2\,
N_{\mathrm{ST}}\over \mathrm{d} t_{\mathrm{vir}}\, \mathrm{d}
M_{\mathrm{vir}}}\, \left|{\mathrm{d}M_{\mathrm{vir}}\over
\mathrm{d}M_{\bullet}}\right|\, \left|{\mathrm{d}M_{\bullet}\over
\mathrm{d}L}\right|~~.\label{eq|previous}
\end{equation}
The above equation corresponds to Eq.~(\ref{eq|QSOLF}), under the
assumptions that the duty-cycle of QSO activity $\Delta
t_{\mathrm{vis}}$ is much shorter than the cosmological time and
that the QSO appears as soon as the host halo virializes, i.e.
$\Delta t_{\mathrm{peak}}\approx 0$.

Wyithe \& Loeb (2003) and Mahmood et al. (2004) estimated the QSO
LFs as a function of cosmic time by assuming: (i) no delay between
virialization and QSO peak luminosity; (ii) a visibility time
$t_{\mathrm{vis}}\approx t_{\mathrm{disc}}$ where the dynamical time
of the galactic disc is $t_{\mathrm{disc}}=0.035\,
R_{\mathrm{vir}}/V_{\mathrm{vir}}\approx 7\times 10^7\,
(1+z)^{-3/2}$ yr; (iii) halo formation rates derived from the
extended Press \& Schechter theory (Lacey \& Cole 1993); (iv) the
$M_{\bullet}-M_{\mathrm{vir}}$ relation deduced from the relations
$M_{\bullet}-\sigma$ and $\sigma-M_{\mathrm{vir}}$ observed by
Ferrarese (2002).

The first and second  assumptions are quite critical at high
redshift. If the time scale of QSO activity is associated to the
dynamical timescale of a galactic disc, one has first to wait for
the disc to be built up, which implies a delay time $\Delta t\ga
\max[t_{\mathrm{cool}}, t_{\mathrm{\mathrm{dyn}}}]$, where both
times are much larger than $t_{\mathrm{disc}}$.

The specification of the DM halo formation rates is also a crucial
step for analytical models. Most of the papers in the literature
adopt the extended Press \& Schechter theory (Bond et al. 1991). In
this paper we have used instead the rates provided by the positive
part of the time derivative of the Sheth \& Tormen mass function.
Note that in both cases the rates are given by the product of the
mass function (Press \& Schechter or Sheth \& Tormen, respectively)
by a factor, weakly dependent on time (Kitayama \& Suto 1996). As a
matter of fact, mass functions derived from numerical simulations
are much better approximated by the Sheth \& Tormen than by the
Press \& Schechter theory; the latter at high redshifts and at the
large halo masses, relevant for QSOs, can be smaller by a factor of
$10$ (e.g., Springel et al. 2005).

As for the last assumptions recalled above, Wyithe \& Loeb (2003)
and Mahmood et al. (2004) assume a $M_{\bullet}-M_{\mathrm{vir}}$
relationship steeper and more rapidly evolving than ours. The
overall result is that, on average, they predict much more massive
BHs at high redshift and for large halo masses (see also the
discussion in Shankar et al. 2006). For example, at $z=6$ and for
$M_{\mathrm{vir}}=10^{13.2}\, M_{\odot}$ we have a BH mass
$M_{\bullet}\approx 2\times 10^9\, M_{\odot}$ while they have
$M_{\bullet}\approx 8\times 10^9\, M_{\odot}$. Recall that we obtain
the relation of Eq.~(\ref{eq|Mbh_Mh}) as a self-consistent output of
the equations in Appendix A, that follow the details of the building
up of the central BH mass.

In summary, the comparison of our results with those of Wyithe \&
Loeb (2003) and Mahmood et al. (2004) is complex, because of quite
different assumptions. Despite of these, similar QSO LFs are
produced. At high redshift their lower $t_{\mathrm{vis}}\la 10^7$ yr
and formation rates of DM halos are compensated by: (i) neglecting
the time needed to build up the galactic structure and (ii) assuming
a relationship $M_{\bullet}-M_{\mathrm{vir}}$ that favors larger BH
masses. The difference emerges quite clearly at redshifts around
$7-8$, where we predict a dramatic drop off of the LF at high
luminosity, because of the delay (see \S~4.1), whereas these authors
obtain number densities larger than ours by a factor of $10$.

Clear differences are also present on the host galaxy side. In our
model the time $\Delta t_{\mathrm{peak}}$ is a very important and
active phase, within which the massive host galaxies are forming
stars at rates around $10^2-10^3\, M_{\odot}$ yr$^{-1}$ (cf.
Fig.~\ref{fig|star&BH}) and the central BH is growing. This phase is
directly explored by sub-mm surveys and by the observations of the
X-ray emission from the growing QSOs detected by Alexander et al.
(2005).

More recently, several authors tried to follow the evolution of both
QSOs and host galaxies using results from numerical simulations, in
which the effects of SN explosions and winds from the central active
nucleus are taken into account (see Hopkins et al. 2006 and
references therein). These authors took an approach complementary to
that presented in this paper. Instead of solving equations
describing the basic physical processes, they simulated many
realizations of collisions between two stable, isolated disk
galaxies endowed with a central supermassive BH. A two-phase
interstellar medium was used to describe star formation and SN
feedback (Springel \& Hernquist 2003). The accretion onto the BHs
was estimated from the local gas density and sound speed, and
limited to the Eddington rate. They assumed that $5\%$ of the
bolometric luminosity derived from accretion is transferred to the
surrounding gas, as thermal energy. They found that during a major
merger the central BHs are fed with enough gas to yield a luminosity
depending exponentially on time. Their approach includes also
estimates of absorption by dust. The energy injected by the QSO in
the interstellar medium is enough to unbind the gas itself on a time
scale $\Delta t_{\mathrm{vis}} \sim 10^9$ yr, within which the
luminosity varies from about $10^{-3}\,L_{\mathrm{peak}}$ to a
maximum $L_{\mathrm{peak}}$, with a weak dependence on
$L_{\mathrm{peak}}$ [cf. Eq.~(7) of Hopkins et al. 2006]. Note that
these simulations do recover the transition from obscured inflow/BH
growth to QSO-driven blowout, as we have simply modeled from the
relevant accretion/feedback equations.

The time scale $\Delta t_{\mathrm{vis}}$ for the growth of the BHs
derived from numerical simulations of low-redshift galaxy mergers at
large halo mass has the same meaning as our time $\Delta
t_{\mathrm{peak}}$ when the maximum luminosity is attained. However,
for large masses and high redshifts the two timescales have quite
different values, $\Delta t_{\mathrm{vis}} \approx 10^9\,
\mathrm{yr} \ga \Delta t_{\mathrm{peak}} \approx 0.2-0.5$ Gyr for
$z_{\mathrm{vir}} \ge 4$. The long time before the peak luminosity
is reached has strong implications at high redshifts. In particular,
$\Delta t_{\mathrm{vis}}\sim t_H$ for $z \ga 4$. Thus the model of
two colliding discs takes too long to form supermassive BHs at high
$z$ (recall that the discs themselves need at least a time
$R_{\mathrm{vir}}/V_{\mathrm{vir}}$ to set up). Therefore the
extension of the model at $z\ga2$ is quite complex, as discussed by
Hopkins et al. (2005) in their \S~3.2. Moreover, it is likely that
several properties inferred from this kind of low-redshift merger
simulations cannot be extrapolated to high redshifts, where the very
idea of binary galaxy mergers likely breaks down. However, it is
interesting to note that at lower redshifts, $z\la 1.5$, the two
times $\Delta t_{\mathrm{vis}}$ and $\Delta t_{\mathrm{peak}}$ are
comparable. Our model is not aimed at following the evolution of
QSOs at $z\la 1$. This later phase is better described by models
that include `interactions' (minor mergers, fly-bys, disk
instabilities, etc.) among galaxies as triggers of nuclear activity
(e.g., Kauffmann \& Haehnelt 2000; Cavaliere \& Vittorini 2000;
Menci et al. 2003).

\begin{figure}[t]
\epsscale{.8}\plotone{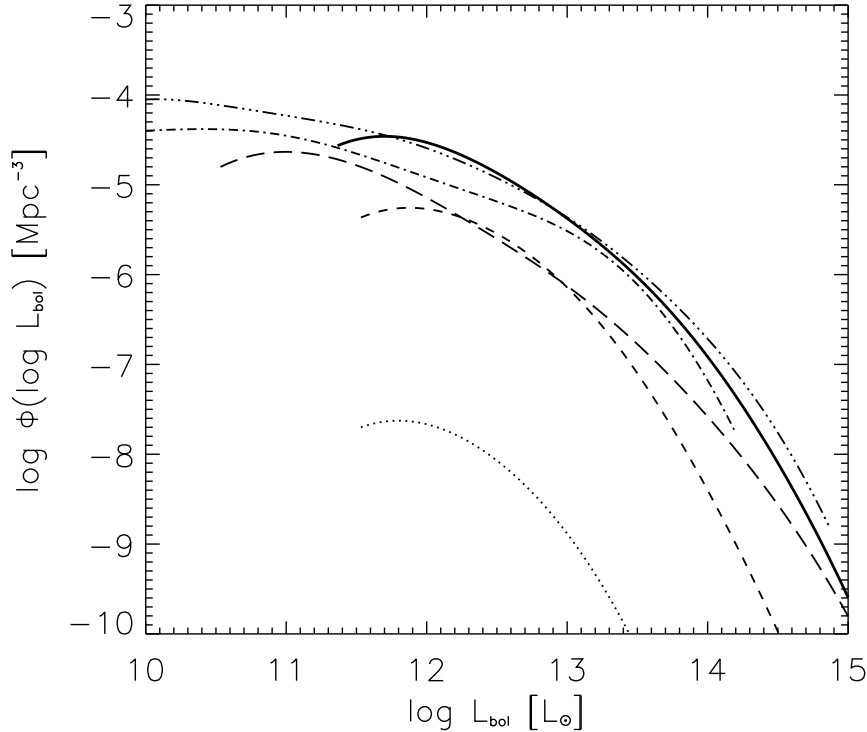}\caption{Bolometric QSO
LFs predicted by our model at redshifts $z\approx 8$
(\textit{dotted} line), $z\approx 6$ (\textit{short dashed} line),
$z\approx 5$ (\textit{long dashed} line), $z\approx 3$
(\textit{solid} line), $z\approx 2$ (\textit{triple dot-dashed}
line), and $z\approx 1.5$ (\textit{dot-dashed} line).}
\label{fig|QSO_LFall}
\end{figure}

\section{Discussion and Conclusions}

In their thoughtful review paper, Brandt \& Hasinger (2005) list
some `key outstanding problems' of AGN astrophysics that need
investigation. These include: the detailed cosmic history of
supermassive BH accretion; the nature of AGN activity in young,
forming galaxies, and the connection between supermassive BH growth
and star formation in sub-mm galaxies.

In this paper we address these issues in the framework of the
anti-hierarchical baryon collapse (ABC) scenario developed by
Granato et al. (2004). We have shown that the condensation of
baryons within DM halos both in stars and in BHs can be described
with a simple physical model, whose main equations are listed in
Appendix A. The model yields the time dependence of the SFR, of the
accretion rate $\dot{M_{\bullet}}$ onto the central BH, of the total
mass in stars $M_{\star}$ and of the final BH mass $M_{\bullet}$,
for any given halo mass and virialization epoch. The GRASIL code
then provides the SED from X-ray to radio bands of evolving stellar
populations (successfully reproducing the epoch-dependent galaxy LFs
in different spectral bands, as well as a variety of relationships
among photometric, dynamical and chemical properties, as shown in
previous papers; see Table~A2), while observationally determined
bolometric corrections allow us to convert the accretion rates onto
supermassive BHs into luminosities in optical and X-ray bands.

In the ABC scenario, the growth rate of BHs is proportional to the
SFR, and the latter is more effectively slowed down by SN feedback
in smaller halos, resulting in a relatively more efficient growth of
more massive BHs, especially at higher redshifts. Both star
formation and BH growth in massive halos are stopped by the feedback
from the active nucleus as soon as it becomes powerful enough to
sweep out the residual interstellar medium.

The BH accretion occurs in host galaxies with very intense star
formation, and is heavily dust obscured for the majority of its
duration; however, most of the final BH mass is gained during the
last $1-2$ $e$-folding times in unobscured conditions, as demanded
e.g. by the Soltan (1982) argument and by the analysis of Yu \&
Tremaine (2002). For massive objects, the growth phase lasts $15-20$
$e$-folding times of Eddington-limited accretion, but the AGNs are
detectable by current hard X-ray surveys only in the last several
$e$-folding times, i.e. have hard X-ray visibility times of $\sim
3\times 10^8\,$yr. Interestingly, Borys et al. (2005) estimate that
the BH masses associated to the X-ray emitting AGNs detected in
sub-mm galaxies are, on average, $\sim 50$ times lower than those
associated to local spheroidal galaxies with similar stellar masses,
as expected if they are about 4 $e$-folding times before the
maximum, for Eddington-limited accretion. At still earlier times,
sub-mm galaxies are expected to host intrinsically weak and highly
obscured nuclei, undetectable with current X-ray telescopes.

The model accounts for the hard X-ray AGN LFs at various redshifts.
A crucial ingredient, to this end, is the rapid quenching of the
accretion rate onto the biggest BHs, predicted by the model as the
consequence of the AGN feedback which sweeps out the interstellar
medium.

The optical (B-band) visibility time is shorter than that for the
more penetrating hard X-rays. The observed epoch-dependent B-band
LFs are accurately reproduced for a $\Delta t_{\mathrm{vis}}$ of
order of the $e$-folding time, indicating that only when the AGN is
approaching its maximum luminosity it can clear up the surrounding
region. The redshift dependence of the space density of optically
bright QSOs is controlled, in the ABC scenario, by two competing
factors. On one side, according to the hierarchical clustering
paradigm, the formation rate of very massive halos hosting them is
increasing rapidly with decreasing redshift. On the other side, the
BH to host halo mass ratio decreases with decreasing redshift, so
that at low redshifts relatively more massive hosts are required for
a given BH mass. But the galactic halo mass function sinks down
exponentially for large halo masses, and is actually cut off at
$M_{\mathrm{vir}}\approx 2\times 10^{13}\,M_{\odot}$. In our model,
the first factor dominates for $z\ga 2.5$ and the second dominates
at lower $z$, thus accounting for the increase in the bright QSO
space density with decreasing $z$ down to $z\approx 2.5$, as well as
for the subsequent decrease.

The redshift dependence of the bolometric QSO LF is illustrated by
Fig.~\ref{fig|QSO_LFall}. It is broadly reminiscent of luminosity
evolution, but with significant deviations from that simple
description particularly at high luminosities. We note in particular
the flattening at high luminosity around $z\approx 5$, borne out by
SDSS data (Fan et al. 2001) and the very sharp drop (by about 2
orders of magnitude) of the QSO LF between $z\approx 6$ and
$z\approx 8$, due to the dearth of massive halos at high redshifts.

As stated in \S~3.1 and 4.1 our model requires a mild variation of
the Eddington ratio $\lambda$ with redshift. This may be a matter of
concern, since the value of $\lambda$ is determined mainly by the
environmental conditions very close to the supermassive BHs. The
main motivation for resorting to super-Eddington accretion at high
$z$ is to account for the observed space density of very luminous
QSOs at $z \ga 5$ (an acceptable fit to lower redshift data can be
obtained keeping $\lambda \approx 1$). Since, as indicated by
Eq.~(\ref{eq|QSOLF}), QSOs seen at the cosmic time $t$ are
associated to halos virializing at the earlier time
$t_{\mathrm{vir}} \la t - \Delta t_{\mathrm{peak}}$, and the density
of such halos drops with increasing $\Delta t_{\mathrm{peak}}$ at
high $z$, to have a sufficient number of massive enough BHs at $z
\ga 5$ we need them to grow faster than implied by Eddington-limited
accretion. The values of $\lambda$ mentioned in \S~3.1 provide the
necessary shortening of $\Delta t_{\mathrm{peak}}$ and also give
high enough luminosities without resorting to too large BH masses.
On the other hand, if we want to preserve the relationships between
the properties of the QSO and of the host galaxy, $\Delta
t_{\mathrm{peak}}$ must be long enough to allow the formation of a
sufficiently massive stellar component.

The redshift dependence of $\lambda$ may thus be seen as a way to
parameterize aspects of the physics of the accretion/emission
process not properly taken into account by our simple model. For
example, a major improvement could be obtained by taking into
account the results of the photon trapping theory (e.g., Blandford
2004). According to the latter, radiative emissions close to
Eddington can be originated by strongly super-Eddington accretion
rates (e.g., those allowed in our model by viscous accretion). This
process seems promising because it may naturally allow short delay
times at high redshift even with an Eddington-limited radiative
output. However, details of the mechanism are still uncertain and
additional work is needed to quantitatively include it in our model.
Non-radiative accretion may also speed up the BH growth. On top of
that, the values of the Eddington ratio could have a substantial
scatter due to different environmental conditions around the
supermassive BHs. As a result, the samples of high redshift QSOs may
be biased towards higher Eddington ratios.

We note that since luminous high-$z$ QSOs are associated to very
massive halos, the model implies that their clustering properties
are similar to those of massive spheroidal galaxies and of the
bright sub-mm galaxies detected by the SCUBA surveys, consistent
with the results by Porciani et al. (2004) and by Croom et al.
(2005). This issue will be investigated in a future paper.

During the relatively short phase when massive galaxies are in the
process of expelling their gas and dust, their nuclei should appear
as powerful X-ray sources, $L_X\ga 10^{45}$ erg s$^{-1}$, but still
somewhat obscured in the optical band; they thus may resemble Type
$2$ QSOs.

Spectroscopic studies, aimed at determining velocity, amount and
chemical abundances of the gas ejected during the QSO phase, would
be extremely informative. In fact, the model predicts that at early
cosmic times large amounts of metal-enriched gas were evacuated from
the galaxy halos to the intergalactic medium.

\begin{acknowledgements}
We warmly thank S. Bressan, A. Celotti, and M. Cirasuolo for helpful
discussions, and the anonymous referee for a careful and perceptive
report, rich of illuminating comments. This work is supported by
grants from ASI, INAF and MIUR.
\end{acknowledgements}

\begin{appendix}

\section{Baryonic physics in our galaxy evolution model}

In this Appendix we provide an overview of the physical model by
Granato et al. (2004), recalling the basic equations and parameters
that control the evolution of the baryonic component.

The baryonic content of a given DM halo with mass $M_{\mathrm{vir}}$
is partitioned in three gaseous phases: a hot diffuse medium with
mass $M_{\mathrm{inf}}$ infalling and/or cooling toward the center;
cold gas with mass $M_{\mathrm{cold}}$ condensing into stars;
low-angular momentum gas with mass $M_{\mathrm{res}}$ stored in a
reservoir around the central supermassive BH, and eventually
viscously accreting onto it. In addition, two condensed phases are
present, namely, stars with a total mass $M_{\star}$ and the BH,
with mass $M_{\bullet}$.

The evolution of the baryonic content is described by the system of
differential equations
\begin{eqnarray}\label{eq|diffeq}
&\nonumber& \dot{M}_{\mathrm{inf}}=
-\dot{M}_{\mathrm{cond}}-\dot{M}_{\mathrm{inf}}^{QSO}~~,\\
&\nonumber&\\
& &\dot{M}_{\mathrm{cold}} = \dot{M}_{\mathrm{cond}}-
\dot{M}_{\star} -\dot{M}_{\mathrm{res}} -
\dot{M}_{\mathrm{cold}}^{SN}-\dot{M}_{\mathrm{cold}}^{QSO}~~,\\
&\nonumber&\\
&\nonumber&\dot{M}_{\mathrm{res}} =
\dot{M}_{\mathrm{inflow}}-\dot{M}_{\bullet}~~,
\end{eqnarray}
where $\dot{M}_{\mathrm{cond}}$ is defined by Eq.~(A2). At the
virialization redshift $z_{\mathrm{vir}}$ we set
$M_{\mathrm{inf}}\approx M_{\mathrm{vir}}/6$ corresponding to the
cosmic baryon to DM ratio and featuring a primordial chemical
composition, $M_{\bullet}\approx M_{\bullet}^{\mathrm{seed}}$,
$M_{\bullet}^{\mathrm{seed}}$ being a BH seed mass originated by
some process in the early universe, and the other baryonic
components to zero.

The hot gas condenses/cools down at the rate
\begin{equation}
\dot{M}_{\mathrm{cond}} = {M_{\mathrm{inf}}\over
\max[t_{\mathrm{cool}}, t_{\mathrm{dyn}}]}~~,
\end{equation}
determined by the dynamical and cooling timescales
$t_{\mathrm{dyn}}$ and $t_{\mathrm{cool}}$ at the virial radius. The
latter includes the appropriate cooling function (Sutherland \&
Dopita 1993) and allows for a clumping factor $\mathcal{C}$ in the
baryonic component. Since at high redshifts major mergers between
massive halos are very frequent, we neglect here the effect of the
angular momentum, since it is lost by dynamical friction through
mergers of mass clouds $M_c$ on time scale $t_{\mathrm{DF}}\approx
0.2\, (\xi/\ln{\xi})\, t_{\mathrm{dyn}}$, where
$\xi=M_{\mathrm{vir}}/M_c$ (see e.g. Mo \& Mao 2004); major mergers
imply $\xi \sim$ a few. In this context it is also worth noticing
that the gas that collapses and cools enough to form stars is only a
fraction of less than $30\%$ of that associated to the virialized
halo (cf. Fig.~\ref{fig|star&BH}).

\begin{deluxetable}{lcllllll}
\tabletypesize{} \tablecaption{Model Parameters} \tablewidth{0pt}
\tablehead{\colhead{} & \colhead{} & \colhead{} &\colhead{} & \colhead{}\\
\colhead{Parameter} & \colhead{} & \colhead{Value} & \colhead{}&
\colhead{Short description}} \startdata
$\mathcal{C}$ & &$7$ & &clumping factor\\
$\epsilon_{SN}$ & &$0.05$ & &strength of SN feedback\\
$\alpha_{RD}$ & & $2.5$ & & strength of radiation drag\\
$\tau_0$ & &1 & &zero-point of optical depth\\
$M_{\bullet}^{\mathrm{seed}}$ & &$10^2\, M_{\odot}$ & & mass of BH seed\\
$k_{\mathrm{accr}}$ & &$10^{-2}$ & & strength of viscous accretion\\
$\lambda$ & & $0.8-4$ & & Eddington ratio$^*$\\
$\eta$ & & $0.15$ & & radiative efficiency\\
$\epsilon_{QSO}$& &$1.3$ & &strength of QSO feedback\\
\enddata
\tablecomments{$^*$We let the maximum allowed Eddington ratio
$\lambda$ to depend on the redshift as: $\lambda=4$ for $z\ga 6$,
$\lambda=3$ for $5\la z\la 6$, $\lambda=1.7$ for $3\la z\la 5$,
$\lambda=1$ for $2\la z\la 3$, and $\lambda=0.8$ for $1.5\la z\la
2$; the empirical fit $\lambda(z)\approx -1.15+0.75\, (1+z)$ works
as well in the redshift range $1.5\la z\la 6$.}
\end{deluxetable}

The cold gas turns into stars at the rate
\begin{equation}
\dot{M}_{\star} =
\int{\mathrm{d}M_{\mathrm{cold}}\over\max[t_{\mathrm{cool}},t_{\mathrm{dyn}}]}~~,
\end{equation}
where now $t_{\mathrm{cool}}$ and $t_{\mathrm{dyn}}$ refer to
cooling and dynamical time, respectively, in the radial mass shell
of mass $\mathrm{d} M_{\mathrm{cold}}$.

The energy feedback from Type-II SNae removes gas from the cold
phase at the rate
\begin{equation}
\dot{M}_{\mathrm{cold}}^{SN} \approx \epsilon_{SN}\,
\left({V_{\mathrm{vir}}\over 500\,\mathrm{km/s}}\right)^{-2}\,
\dot{M}_{\star}~~,\label{eq|snfeed}
\end{equation}
where the efficiency $\epsilon_{SN}$ is a parameter of the model.
The amount of cold mass removed is proportional to the number of SN
explosions (assuming an average energy of around $10^{51}$ ergs per
SN) hence to the SFR, and inversely proportional to the depth of the
halo potential well.

The stellar radiation drag (see Kawakatu \& Umemura 2002) triggers
the inflow of cold gas into a reservoir of low-angular momentum
around the central supermassive BH, at the rate
\begin{equation}
\dot{M}_{\mathrm{inflow}} \approx \alpha_{RD}\times 10^{-3} \,
\dot{M}_{\star}\, (1-e^{-\tau})~~ M_{\odot}~ \mathrm{yr}^{-1}~~,
\end{equation}
where the effective optical depth is given by
\begin{equation}
\tau = \tau_0\, \left({Z\over Z_{\odot}}\right)\,
\left({M_{\mathrm{cold}}\over 10^{12}\, M_{\odot}}\right)\,
\left({M_{\mathrm{vir}}\over 10^{13}\, M_{\odot}}\right)^{-2/3}~~.
\end{equation}
The metallicity $Z(t)$ is computed from the code self-consistently,
while the coefficients $\alpha_{RD}$ and $\tau_0$ are parameters
(see Granato et al. (2004) for further details).

The viscous time $t_{\mathrm{visc}}=\mathrm{Re}_{\mathrm{crit}}\,
t_{\mathrm{dyn}}^{\mathrm{BH+res}}$ for the accretion from the
reservoir to the BH is related to the dynamical time
$t_{\mathrm{dyn}}^{\mathrm{BH+res}}$ of the system BH $+$ reservoir,
and to the critical Reynolds number $\mathrm{Re}_{\mathrm{crit}}\sim
10^2-10^3$ (see Granato et al. 2004 for details). Thus the gas in
the reservoir accretes onto the supermassive BH at the rate
\begin{equation}
\dot{M}_{\bullet}^{\mathrm{visc}}\approx 5\times 10^3 \,
k_{\mathrm{accr}}\, \left(\frac{V_{\mathrm{vir}}}{500~ \mathrm{km~
s}^{-1}}\right)^3\, \left({M_{\mathrm{res}}\over
M_{\bullet}}\right)^{3/2}\,\left(1+{M_{\mathrm{\bullet}}\over
M_{\mathrm{res}}}\right)^{1/2}~~ M_{\odot}~ \mathrm{yr}^{-1}~~.
\end{equation}
Again, the coefficient $k_{\mathrm{accr}}$ is a model parameter. For
the plausible range of $10^{-4}\la k_{\mathrm{accr}}\la 10^{-2}$,
the accretion rate $\dot{M}_{\bullet}^{\mathrm{visc}}$ easily
exceeds the Eddington rate for high redshifts and massive halos. In
this situation we may have: (i) very rapid growth of the central BH
in a dynamical timescale, followed by a slower accretion when less
mass is available in the reservoir; (ii) BH growth limited by
radiation pressure, followed by strong mass outflows $\dot{M}_w \sim
\dot{M}_{\mathrm{Edd}}$ (e.g., Small \& Blandford 1992; King \&
Pounds 2003; Begelmann 2004). We select the latter option and the
actual accretion rate is set to
\begin{equation}
\dot{M}_{\bullet} =\min[\dot{M}_{\bullet}^{\mathrm{visc}}, \lambda\,
\dot{M}_{\bullet}^{\mathrm{Edd}}]~,
\end{equation}
where
\begin{equation}
\dot{M}_{\bullet}^{\mathrm{Edd}}\approx 1.3\,
\left({M_{\bullet}\over 10^8\, M_{\odot}}\right)~~ M_{\odot}~
\mathrm{yr}^{-1}~~
\end{equation}
is the Eddington rate and $\lambda$ is the maximum allowed Eddington
ratio. We let the maximum allowed Eddington ratio $\lambda$ to
depend on the redshift as: $\lambda=4$ for $z\ga 6$, $\lambda=3$ for
$5\la z\la 6$, $\lambda=1.7$ for $3\la z\la 5$, $\lambda=1$ for
$2\la z\la 3$, and $\lambda=0.8$ for $1.5\la z\la 2$. The empirical
fit $\lambda(z)\approx -1.15+0.75\, (1+z)$ works as well in the
redshift range $1.5\la z\la 6$, see also Table~A1.

In the presence of Eddington-limited or super-Eddington accretion,
QSO-driven outflows are expected in the form of winds that can form
just above the disc by a combination of radiation and gas pressure.
Large mass outflows have been confirmed by X-ray observations of BAL
QSOs (Brandt \& Gallagher 2000; Chartas et al. 2003). Following the
model by Murray et al. (1995) the asymptotic speed and the mass
outflow rates are estimated as $v_w\approx 0.06\, c\, (r_w/10^{16}\,
\mathrm{cm})^{-1/2}\, (\dot{M}_{\bullet}/M_{\odot}\,
\mathrm{yr}^{-1})^{1/2}$ and $\dot{M}_w\approx 2.4\, f_c\,
N_{22}\,\, (r_w/10^{16}\, \mathrm{cm})^{1/2}\,
(\dot{M}_{\bullet}/M_{\odot}\, \mathrm{yr}^{-1})^{1/2}\,$
$M_{\odot}$ yr$^{-1}$, respectively; here $c$ is the speed of light,
$f_c$ is the covering factor, $N_{22}$ is the hydrogen column
density normalized to $10^{22}$ cm$^{-2}$, and $r_w$ is the
launching radius. Then the kinetic luminosity of the outflow reads
\begin{equation}
L_K=\frac{1}{2}\, \dot{M}_{w}\, v_{w}^2 \approx 2.5\times 10^{44}\,
\left(\frac{f_c}{0.1}\right)\,\left(\frac
{N_{22}}{10}\right)\,\left(\frac{r_w}{10^{16}\,
\mathrm{cm}}\right)^{-1/2}\, \left({\dot{M}_{\bullet}\over
M_{\odot}/ \mathrm{yr}}\right)^{3/2}\,\hbox{erg}\,\hbox{s}^{-1}.
\end{equation}
Eventually, QSO outflows remove gas from both the hot and the cold
phases at the rates
\begin{equation}
\dot{M}_{\mathrm{inf, cold}}^{QSO} = 2\times 10^3\, \epsilon_{QSO}\,
\left({V_{\mathrm{vir}}\over 500\,\mathrm{km/s}}\right)^{-2}\,
\left({\dot{M}_{\bullet}\over M_{\odot}/ \mathrm{yr}}\right)^{3/2}\,
{M_{\mathrm{inf, cold}}\over M_{\mathrm{cold}}+M_{\mathrm{inf}}}~~
M_{\odot}~ \mathrm{yr}^{-1},
\end{equation}
with $\epsilon_{QSO}=(f_h/0.5)\, (f_c/0.1)\, (N_{22}/10)$, where
$f_h$ is the fraction of the kinetic luminosity that couples to the
gas. Note that most of the power $L_h=f_h\, L_K$ is transferred to
the gas near the peak of the BH accretion rate curve. Such a power
is only a small fraction of the Eddington luminosity
$(L_h/L_{\mathrm{Edd}})\approx 1.4 \times 10^{-2}\, \epsilon_{QSO}\,
(M_{\bullet}/10^8\, M_{\odot})$.

\begin{deluxetable}{lll}
\tabletypesize{} \tablecaption{Overview of Model Results}
\tablewidth{0pt} \tablehead{\colhead{Property or Statistics} &
\colhead{External Inputs$^{\star}$} & \colhead{Reference}}\startdata
\\
\sidehead{\textit{QSOs and BHs}}
$M_{\bullet}-M_{\mathrm{vir}}$ & no add & Fig.~2 (left)\\
$\Delta t_{\mathrm{peak}}-M_{\mathrm{vir}}$ & no add & Fig.~2 (right)\\
Opt. LFs $(z)$ & $\Delta t_{\mathrm{vis}}\approx 5\times 10^7$ yr, $\Delta \log{M_{\bullet}}\approx 0.3$ & Figs.~3-6\\
X-ray LFs $(z)$ & $\Delta t_{\mathrm{vis}}\approx 3\times 10^8$ yr & Figs.~5-6 (right)\\
BH mass function & no add & Fig.~7 \\
$M_{\bullet}-\sigma$ & no add & Fig.~6 of G04\\
\\
\sidehead{\textit{Spheroidal Galaxies}}
VDF & $\sigma/V_{\mathrm{vir}}$ & Fig.~1 of C05\\
Faber-Jackson & $\sigma/V_{\mathrm{vir}}$ & Fig.~2 of C05\\
LF $(z)$ & IMF, SEDs (GRASIL) & Fig.~10 of G04  \\
Metal abundances & IMF, chemical yields & Fig.~9 of G04 \\
K-band counts & IMF, dust modeling (GRASIL) & Fig.~1 of S05\\
$850\,\mu$m counts & IMF, dust modeling (GRASIL) & Fig.~12 of G04\\
EROs & IMF, dust modeling (GRASIL) & Fig.~6 of S05 \\
\\
\enddata
\tablecomments{$^\star$ Input quantities not included in Table~A1.
G04: Granato et al. (2004), C05: Cirasuolo et al. (2005), S05: Silva
et al. (2005).}
\end{deluxetable}

The basic outputs of the model are the evolution of the SFR
$\dot{M}_{\star}(t)$ and the BH accretion rate
$\dot{M}_{\bullet}(t)$ as a function of the galactic age $t$ for any
given value of the halo mass $M_{\mathrm{vir}}$ and of the
virialization redshift $z_{\mathrm{vir}}$. Once the star formation
history of a galaxy has been computed, its luminosity in any chosen
band is obtained as a function of $t$ from the GRASIL code, which
yields the chemical and the spectro-photometric evolution from the
radio to the X-ray band, allowing for the effect of dust absorption
and reradiation.

\end{appendix}

\end{document}